\documentclass[%
  aps, reprint, a4paper, twocolumn, superscriptaddress, amsmath, amssymb, floatfix
]{revtex4-2}
\usepackage{amsthm}
\usepackage{graphicx}
\usepackage{epstopdf}
\usepackage[table]{xcolor}
\usepackage{braket}
\usepackage[colorlinks,bookmarksopen=false]{hyperref}
\usepackage[caption=false]{subfig}
\usepackage{bbm}
\usepackage{makecell}
\usepackage{multirow}
\usepackage{lipsum, blindtext}
\usepackage{xr}
\usepackage{MnSymbol}
\usepackage{siunitx}
\usepackage[normalem]{ulem}
\usepackage{cancel}
\usepackage{threeparttable}

\hypersetup{%
  pdfstartview={FitW},
  linkcolor=blue,
  citecolor=blue,
  filecolor=black,
  menucolor=black,
  urlcolor =blue
}

\newcommand{\im}{\ensuremath{\textup{i}}}

\newcommand{\rp}{\ensuremath{\mathfrak{Re}}}
\newcommand{\ip}{\ensuremath{\mathfrak{Im}}}

\newcommand{\op}[1]{\ensuremath{\mathsf{#1}}}

\newcommand{\error}{\varepsilon}
\newcommand{\pulse}{\mathcal{E}}

\newtheorem*{mydef*}{Definition}
\newtheorem*{myprop*}{Proposition}

\newtheorem*{mythm*}{Theorem}

\newcommand{\dd}{\mathrm{d}}

\newcommand{\rmdiag}{\mathrm{diag}}
\newcommand{\rmbound}{\mathrm{bound}}
\newcommand{\rmphys}{\mathrm{phys}}

\newcommand{\rmlit}{\mathrm{lit}}

\newcommand{\rmrot}{\mathrm{rot}}
\newcommand{\rmref}{\mathrm{ref}}
\newcommand{\rmint}{\mathrm{int}}

\newcommand{\rmryd}{\mathrm{Ryd}}

\newcommand{\rmopt}{\mathrm{opt}}
\newcommand{\trgt}{\mathrm{trgt}}
\newcommand{\rmin}{\mathrm{in}}
\newcommand{\rmout}{\mathrm{out}}
\newcommand{\rmre}{\mathrm{re}}
\newcommand{\rmim}{\mathrm{im}}

\newcommand{\rmx}{\mathrm{x}}

\newcommand{\rmz}{\mathrm{z}}
\newcommand{\rmmax}{\mathrm{max}}
\newcommand{\rmmin}{\mathrm{min}}

\newcommand{\rmqsl}{\mathrm{QSL}}

\newcommand{\rma}{\mathrm{a}}

\newcommand{\rmc}{\mathrm{c}}

\newcommand{\Omd}{\Omega_{\downarrow}}
\newcommand{\Omu}{\Omega_{\uparrow}}
\newcommand{\Omdn}{\Omega_{\downarrow,n}}
\newcommand{\Omun}{\Omega_{\uparrow,n}}
\newcommand{\phid}{\varphi_{\downarrow}}
\newcommand{\phiu}{\varphi_{\uparrow}}
\newcommand{\phidn}{\varphi_{\downarrow,n}}
\newcommand{\phiun}{\varphi_{\uparrow,n}}

\begin{document}

\title{%
    Comparing planar quantum computing platforms at the quantum speed limit
}

\author{Daniel Basilewitsch}
\email{daniel.basilewitsch@uibk.ac.at}
\affiliation{%
  Institute for Theoretical Physics, University of Innsbruck, A-6020 Innsbruck,
  Austria
}

\author{Clemens Dlaska}
\affiliation{%
  Institute for Theoretical Physics, University of Innsbruck, A-6020 Innsbruck,
  Austria
}

\author{Wolfgang Lechner}
\affiliation{%
  Institute for Theoretical Physics, University of Innsbruck, A-6020 Innsbruck,
  Austria
}
\affiliation{%
  Parity Quantum Computing GmbH, A-6020 Innsbruck, Austria
}
\date{\today}

\begin{abstract}
  An important aspect that strongly impacts the experimental feasibility of
  quantum circuits is the ratio of gate times and typical error time scales.
  Algorithms with circuit depths that significantly exceed the error time scales
  will result in faulty quantum states and error correction is inevitable. We
  present a comparison of the theoretical minimal gate time, i.e., the quantum
  speed limit (QSL), for realistic two- and multi-qubit gate implementations in
  neutral atoms and superconducting qubits. Subsequent to finding the QSLs for
  individual gates by means of optimal control theory we use them to quantify
  the circuit QSL of the quantum Fourier transform and the quantum approximate
  optimization algorithm. In particular, we analyze these quantum algorithms in
  terms of circuit run times and gate counts both in the standard gate model and
  the parity mapping. We find that neutral atom and superconducting qubit
  platforms show comparable weighted circuit QSLs with respect to the system
  size.
\end{abstract}

\maketitle

\section{Introduction}
Quantum computers promise to solve computational problems that are deemed hard
or even intractable for classical computers. Their potential applications
include prime-factoring of large integers~\cite{SIAM.26.1484}, quantum
simulation~\cite{Daley2022}, quantum chemistry~\cite{Cao2019}, combinatorial
optimization~\cite{Farhi2014}, and even problems in finance~\cite{Orus2019}.
Currently, quantum computing is in the so-called noisy intermediate-scale
quantum (NISQ) era~\cite{PreskillQ2018}, characterized by imperfect qubit
control, and qubit numbers that prohibit quantum error
correction~\cite{RMP.87.307} for relevant problem sizes. Nevertheless, recent
proof-of-principle experiments~\cite{Arute2019, PRL.127.180501, Zhong2020,
PRL.127.180502} demonstrated that a computational quantum advantage over
classical computers can be reached already with NISQ hardware. However, it
remains a crucial challenge to go beyond the proof-of-principle stage, i.e., to
demonstrate a quantum advantage for practically relevant computational tasks on
resource-limited present-day devices. 
\par
To reach a practical quantum advantage regime~\cite{Daley2022} in NISQ-era
digital quantum computing it is of crucial importance to execute quantum
algorithms as efficiently as possible in order to minimize the time for noise
mechanisms to impair the quantum information processing. This effectively makes
a minimization of the quantum algorithm run times and gate counts desirable ---
a task that can be addressed in various ways. One option is to find an
algorithm's optimal circuit representation, i.e., a circuit requiring a minimal
circuit depth together with a minimal gate count for a given set of available
gates. Quantum circuit optimization has, e.g., been done
heuristically~\cite{Amy2014, Heyfron2018} or by machine learning
techniques~\cite{PRA.98.032309, Foesel2021} and several open-source packages are
readily available~\cite{Sivarajah2020, PRA.102.022406, Nagarajan2021}.
\par
Another option for minimizing the algorithm run time is to minimize the time for
each elemental quantum gate of a given quantum circuit. While protocols for
fast, high-fidelity quantum gates are nowadays routinely available and
implemented on all major quantum computing platforms like neutral
atoms~\cite{PRL.85.2208, PRL.123.170503}, superconducting
circuits~\cite{PRL.103.110501, Arute2019} or trapped ions~\cite{PRL.91.157901,
Schfer2018}, the ideal would be to execute every quantum gate at its quantum
speed limit (QSL). In general, the QSL denotes the shortest time needed to
accomplish a given task~\cite{Deffner2017}. It constitutes a fundamental limit
in time and depends on the system under consideration, i.e., its Hamiltonian and
the control knobs available to steer the dynamics.

Here, we determine the QSLs of quantum gates for two major quantum computing
platforms that allow for two-dimensional (2D) qubit arrangements --- neutral
atoms and superconducting circuits~\footnote{Due to the restriction to 2D
architectures, we do not consider trapped ion platforms in this study}. Our
study reveals how close the current experimental gate protocols are in
comparison to their QSLs and thus exemplifies what can theoretically still be
gained from further speeding up gate protocols. Moreover, provided that every
gate could be experimentally realized at the QSL, our analysis gives an estimate
of how many gates can be executed realistically before decoherence takes over
and renders longer quantum circuits practically infeasible. To this end, we
consider two prototypical quantum computing algorithms: (i) the quantum Fourier
transform (QFT), required for Shor's algorithm for integer
factorization~\cite{SIAM.26.1484}, and (ii) the quantum approximate optimization
algorithm (QAOA), used to solve combinatorial optimization
problems~\cite{Farhi2014}. Considering standard NISQ devices for both neutral
atoms and superconducting circuits with qubits arranged in a 2D grid
architecture with only nearest-neighbor connectivity, we calculate the circuit
run times with gates at the QSL for both algorithms. This allows for a direct
comparison of both platforms in terms of the maximal problem sizes that should
currently be feasible on their NISQ representatives.

A common challenge arising in 2D platforms with nearest-neighbor connectivity is
the requirement to perform gates between non-neighboring qubits. In the standard
gate model (SGM), such gates can be replaced by sequences of universal single-
and two-qubit gates using the available local connectivity. However, this comes
at the price of increasing the circuit depths and gate counts. As an alternative
to the SGM, we also examine circuit representations using the so-called parity
mapping (PM). In brevity, the PM for quantum computing~\cite{PRL.129.180503} and
quantum optimization~\cite{Lechner2015, Ender2023} is a problem-independent
hardware blueprint that only requires nearest-neighbor connectivity at the cost
of increased qubit numbers. Since for QAOA circuits in the PM it is beneficial
to use local three- and four-qubit gates~\cite{Lechner2020, Ender2022}, we also
determine their QSLs on both platforms.

It should be noted that this work focuses entirely on the determination of the
QSLs for various quantum gates and how to turn these into a fair comparison of
circuit run times across platforms. A more detailed discussion of gate protocols
for specific quantum gates or platforms can, e.g., be found in
Refs.~\cite{PRR.4.033019, Pelegr2022, Jandura2022} for neutral atoms or in
Refs.~\cite{PRL.112.240504, Goerz2017, Werninghaus2021} for superconducting
circuits. Moreover, it should be noted that we only consider the gate times,
respectively circuit run times, as well as the number of gates as indicators for
the feasibility of quantum circuits. While both quantities are doubtlessly
important, they are by no means the only quantities impacting a circuit's
feasibility. A holistic figure of merit assessing a circuit's feasibility would
also need to account for state preparation and measurement errors and various
other error sources. 

The paper is organized as follows. In Sec.~\ref{sec:main} we present our main
result, i.e., an overview of circuit run times when using gate times from the
literature and gate times at the QSL --- evaluated both for neutral atoms and
superconducting circuits as well as for circuits in the SGM and the PM. In
Sec.~\ref{sec:model} we then present the details of our numerical model.
Section~\ref{sec:oct} introduces the basic notion of QSLs and how quantum
optimal control theory (OCT) can be used for determining QSLs. A detailed
discussion of QSLs as well as which combinations of available control fields
allow to reach the QSLs, is given in Sec.~\ref{sec:bench}.
Section~\ref{sec:conclusions} concludes.

\section{Main result: quantum algorithms at the quantum speed limit}
\label{sec:main}
In this section, we compare the run times of QFT and QAOA quantum circuits using
gate set implementations available on neutral atom and superconducting circuit
hardware. To do so we consider two scenarios. In the first scenario, we use
literature values for gate times of state-of-the-art gate implementations of the
minimal universal gate set (referred to as ``standard gate set'' (SGS) from now
on) native to each platform, which thus represents the canonical way of
converting quantum algorithms into executable quantum circuits. In the second
scenario, we use an extended set of gates available at each platform with gate
times at the QSL (referred to as ``QSL gate set'' (QGS) from now on), which thus
yields the current fundamental limits in circuit run times. We describe both
gate sets in Sec.~\ref{subsec:gate sets} and use them to analyze circuit run
times of QFT and QAOA quantum circuits both in the SGM in
Sec.~\ref{subsec:standard} and the PM in Sec.~\ref{subsec:parity}. A brief
review of the basic concepts and quantum circuits of a QFT and a single QAOA
step within the SGM and the PM is given in Appendix~\ref{app:circs}.

\begin{table}[tb!]
  \renewcommand*{\arraystretch}{1.2}
  \begin{threeparttable}
    \begin{tabular*}{\columnwidth}{@{\extracolsep{\stretch{1}}}ccccc}
      \hline 
      &
      \multicolumn{2}{c}{%
        \makecell{neutral \\ atoms}
      }
      &
      \multicolumn{2}{c}{%
        \makecell{superconducting \\ circuits}
      }
      \\ \hline 
      \multirow{3}{*}{%
        \makecell{standard \\ gate set \\ (SGS)}
      }
      & gate & time (ns) & gate & time (ns)
      \\ \cline{2-3} \cline{4-5}
      &
      local         & $\sim 1000$~\cite{PRL.123.170503} &
      local         & $       25$~\cite{datasheet}
      \\
      &
      CZ            & $ 350$~\cite{PRL.123.170503} &
      Sycamore      & $  12$~\cite{datasheet}
      \\ \hline 
      \multirow{7}{*}{%
        \makecell{QSL \\ gate set \\ (QGS)}
      }
      & gate & time (ns) & gate & time (ns)
      \\ \cline{2-3} \cline{4-5}
      &
      local         & $\sim 1000$~\cite{PRL.123.170503} &
      local         & $       25$~\cite{datasheet}
      \\
      &
      CNOT$^{\rma}$ & $ 300$ &
      CNOT          & $  14$
      \\
      &
      CZ            & $ 350$ &
      CZ            & $  10$
      \\
      &
      SWAP          & $ 400$ &
      SWAP          & $  12$
      \\
      &
      ZZZ           & $ 600$ &
      ZZZ           & $  24$
      \\
      &
      ZZZZ          & $ 600$ &
      ZZZZ          & $  80$
      \\ \hline 

    \end{tabular*}
    \begin{tablenotes}\footnotesize
      \item[a] The CNOT gate is listed for completeness but not used in any
        circuits for neutral atoms, see Sec.~\ref{sec:bench} for details.
    \end{tablenotes}
  \end{threeparttable}
  \caption{%
    Overview of different gate sets and corresponding gate times on neutral
    atoms and superconducting circuits. While the row ``standard gate set''
    (SGS) represents typical gates and times used on the respective platform,
    the multi-qubit gates in the row named ``QSL gate set'' (QGS) correspond to
    an extended gate set with gate times at the QSL. 
  }
  \label{tab:gatesets}
\end{table}

\begin{figure*}[tb!]
  \centering
  \includegraphics[width=1.0\textwidth]{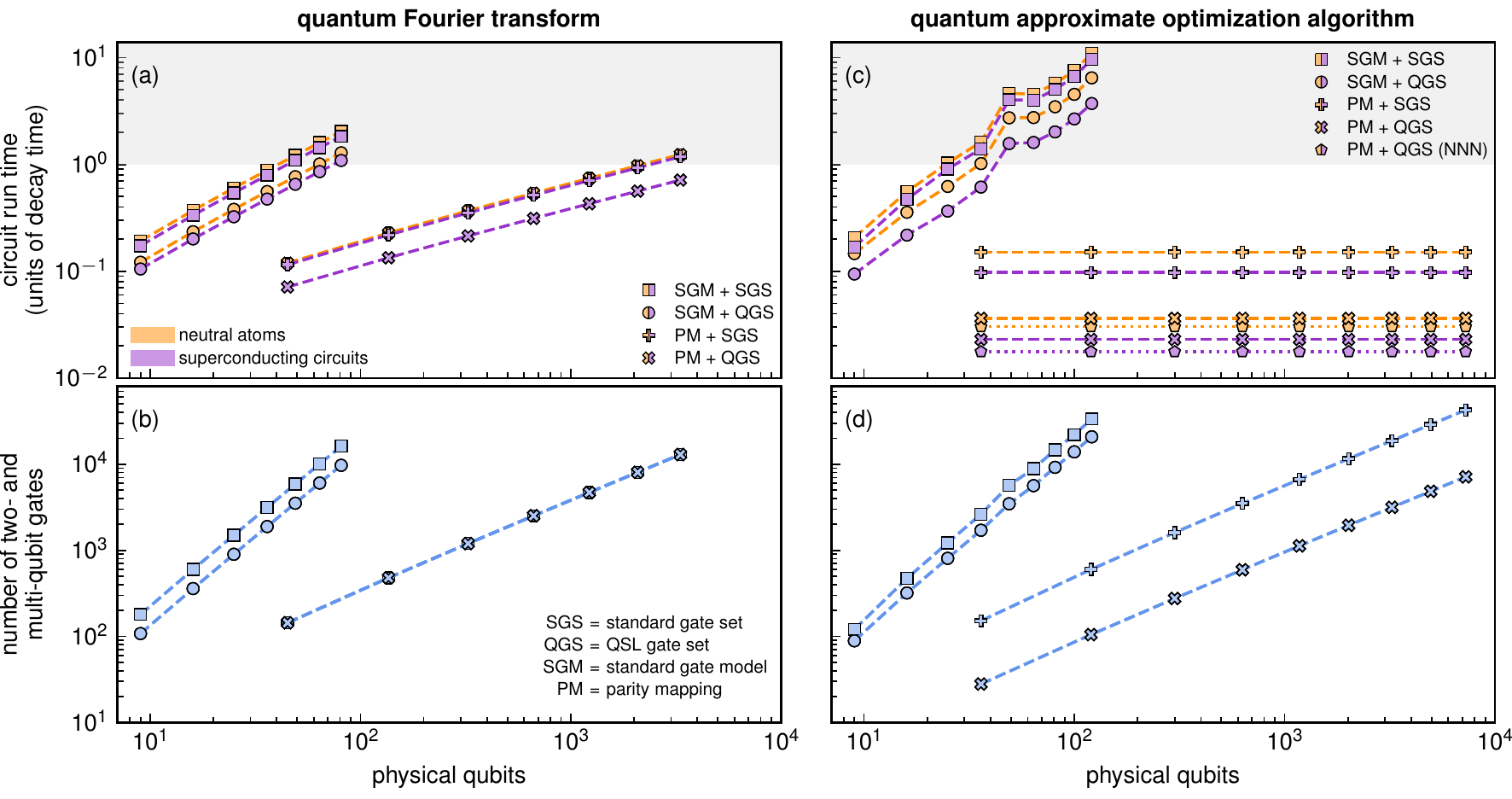}
  \caption{%
    Comparison of the circuit run times and gate counts for the quantum Fourier
    transform (QFT) and quantum approximate optimization algorithm (QAOA)
    between neutral atoms and superconducting circuits. The left (right) column
    corresponds to the QFT (QAOA) and the orange (purple) marker correspond to
    neutral atoms (superconducting circuits). The circuit run times for both
    algorithms and various problem instances of different size, $N=9, 16, \dots,
    81$ qubits for the QFT and $N=9, 16, \dots, 121$ qubits for the QAOA, are
    given in panels (a) and (c), respectively. The numbers of two- and
    multi-qubit gates in the corresponding circuits are given in panels (b) and
    (d), respectively. The data for the squares [circles] is obtained using the
    standard gate model (SGM) using the gates and times from the standard gate
    set (SGS) [QSL gate set (QGS)], cf.\ Table~\ref{tab:gatesets}. In contrast,
    the plus signs [crosses] represent the results when the same algorithms are
    realized in the parity mapping (PM) using the SGS [QGS]. The pentagons
    correspond to an altered 2D architectures allowing for next-to-nearest
    neighbor (NNN) coupling between qubits. The run time for each circuit is
    given in units of typical platform-specific error times. The gray area in
    panels (a) and (c) indicates where the circuit run times exceeds the error
    times.
  }
  \label{fig:comp}
\end{figure*}

\begin{figure}[tb!]
  \centering
  \includegraphics{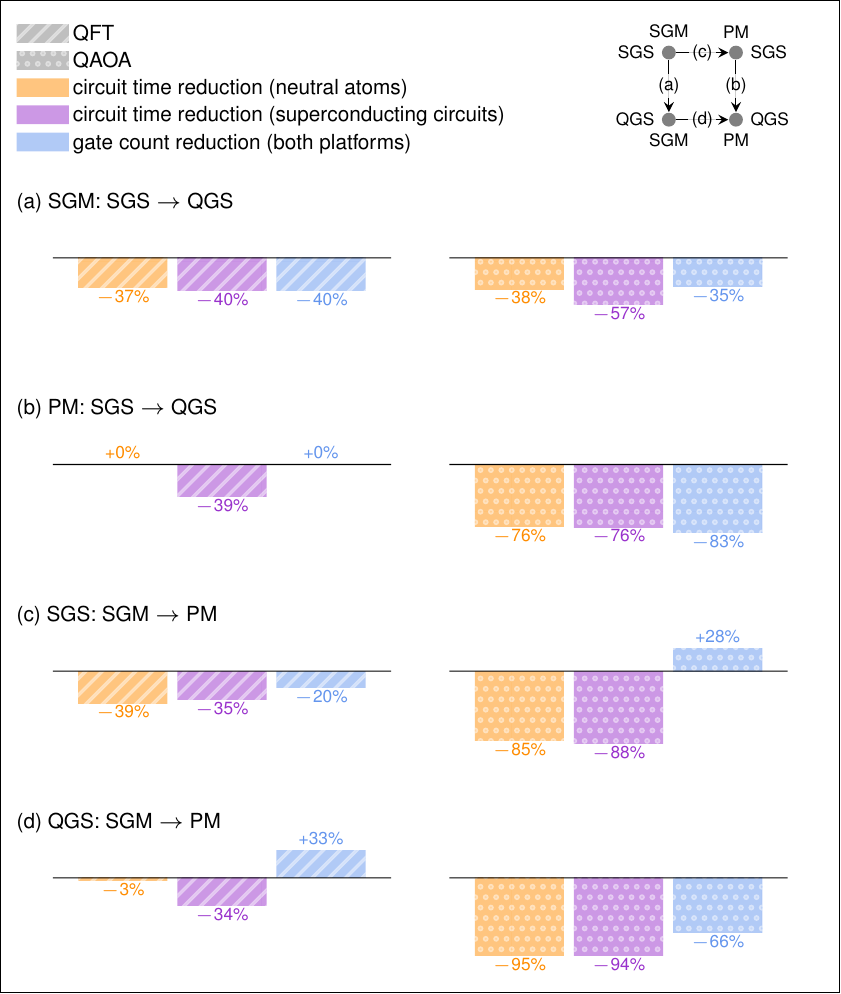}
  \caption{%
    Reduction in weighted circuit run times and (two- and multi-qubit) gate
    counts for the scenarios and data shown in Fig.~\ref{fig:comp}. In panels
    (a) and (b) the circuits in the standard gate model (SGM) and parity mapping
    (PM) are compared when using the ``QSL gate set'' (QGS) instead of the
    ``standard gate set'' (SGS). In panels (c) and (d) the circuits in the SGS
    and QGS are compared when using the PM instead of the SGM. All presented
    numbers are the average over problem instances of different sizes, i.e.,
    different numbers of qubits.
  }
  \label{fig:stats}
\end{figure}

\subsection{Standard and QSL gate sets}
\label{subsec:gate sets}
In the SGM, each quantum algorithm is converted into quantum circuits using
gates from a universal set of quantum gates. Such a universal set typically
contains all the single-qubit gates and at least one entangling two-qubit gate,
e.g., the CNOT gate~\cite{PRA.52.3457}. The row named ``standard gate set'' in
Table~\ref{tab:gatesets} summarizes the native universal gate sets and their
typical gate times for both platforms at comparable gate fidelities. For neutral
atoms, we consider the controlled-Z gate, $\mathrm{CZ} = \rmdiag\{1,1,1,-1\}$,
as the typical entangling two-qubit gate. This is a common choice for neutral
atoms~\cite{PRL.123.170503} as it has been successfully used for implementing
quantum algorithms~\cite{Bluvstein2022, Graham2022}. For superconducting
circuits, we have chosen the iSWAP-like Sycamore gate as an entangling two-qubit
gate, motivated by its short gate time and successful usage in recent quantum
advantage experiments with quantum processors based on tunable
couplers~\cite{Arute2019, PRL.127.180501}. 
\par
In addition, we consider an extended, platform-independent set of quantum gates
operated at the QSL (see ``QSL gate set'' in Table~\ref{tab:gatesets}). This set
consists of all single-qubit gates as well as several multi-qubit gates
implementable on both platforms: CZ, CNOT, SWAP, ZZZ, and ZZZZ. The availability
of a wider range of gates allows for more flexibility in finding circuit
representations with fewer gates and shorter circuit run times. In addition, we
also assume that every gate in this set is executed at the QSL, which allows for
another speed-up of circuit run times. The run times obtained for the QGS should
therefore be viewed as an estimate for the QSL of the circuit itself, i.e., the
circuit QSL --- provided that no other, potentially faster and/or better suited
gates are available~\footnote{Note that in practice, gate sets are typically
tailored to match the quantum algorithm in question. However, this goes beyond
the scope of this work}. Note that the details regarding the method to determine
the QSLs and their results are presented in Secs.~\ref{sec:oct} and
\ref{sec:bench}. The row ``QSL gate set'' in Table~\ref{tab:gatesets} summarizes
the gates of the extended set and lists the QSL times for both platforms.
\par
Table~\ref{tab:gatesets} indicates that absolute circuit run times for neutral
atoms will be longer compared to those for superconducting circuits since their
elemental gate times differ by more than an order of magnitude. However, in
order to ensure a fair comparison of circuit run times the absolute gate times
need to be weighted by the finite coherence time and lifetime of qubits and
other levels involved in the gate mechanisms.
\par
For neutral atoms, we take both the coherence time of the qubit states, given by
the dephasing time $T_{2}^{*} = \SI{4}{\milli\second}$~\cite{Bluvstein2022}, and
the lifetime of the Rydberg state, $T_{\rmryd}
= \SI{150}{\micro\second}$~\cite{Bluvstein2022}, as typical error time scales
against which we compare circuit run times. We take the former as reference for
single-qubit gates, since they don't occupy the Rydberg level~\footnote{Assuming
that the qubit is encoded in the atomic ground state manifold, such as hyperfine
states.},and the latter for two-qubit gates, where controlled transitions via
Rydberg levels constitute the primary gate mechanism~\footnote{Note that we do
not consider the lifetime of the intermediate levels between qubit and Rydberg
states}. For superconducting circuits, we take their intrinsic $T_{1}$ time,
$T_{1} = \SI{15}{\micro\second}$~\cite{Arute2019}, as typical error time scale
since it applies for both single- and two-qubit gates. However, note that these
choices are rather conservative. For neutral atoms, any two-qubit gate dynamics
will naturally also involve the qubit levels, which have much longer coherence
times. Weighting two-qubit gates exclusively by the Rydberg lifetime thus
overestimates the lifetime-induced error probability. For superconducting
circuits, longer $T_{1}$ times up to $\SI{500}{\micro\second}$ have already been
reported~\cite{Wang2022}.

\subsection{Circuit times in the standard gate model}
\label{subsec:standard}
\par
In the following, we calculate the circuit run times for QFTs and single QAOA
steps in the SGM using gates from both the SGS and the QGS. As already outlined
above, we assume both hardware platforms to consist of qubits arranged in 2D
arrays with only nearest-neighbor connectivity (see Sec.~\ref{sec:model} for
details regarding the model). This requires replacing all gates between
non-connected qubits with gate sequences between physically connected qubits. 

\subsubsection{Standard gate set circuits}
Using gates from the SGS, the circuit's gate sequences consist of single-qubit
gates plus CZ gates in case of neutral atoms and single-qubit gates plus
Sycamore gates in case of superconducting circuits
(cf.~Table~\ref{tab:gatesets}). Since the circuit for the QFT does by
construction not require any gate between non-neighboring
qubits~\cite{Holmes2020}, we only replace the controlled-phase gate and SWAP
gate by gate sequences from the SGS. The situation changes for the QAOA
circuits, as it assumes an all-to-all connected architecture and finding
a circuit representation with minimal depth while only requiring
nearest-neighbor connectivity is likely NP-hard~\cite{Harrigan2021}. As
a remedy, we use the open-source \emph{pytket} compiler~\cite{Sivarajah2020} to
translate the QAOA's bare quantum circuits to executable quantum circuits that
are in agreement with the hardware's nearest-neighbor connectivity and SGSs. We
furthermore use the compiler's quantum circuit optimization features to optimize
the circuits in order to minimize gate counts and circuit depths.

The squares in Fig.~\ref{fig:comp} (a) and (c) show the resulting run times for
quantum circuits corresponding to QFTs and QAOA steps, respectively, when
executed on neutral atoms (orange) and superconducting circuits (purple). Within
each line, the size of the problem instance increases from the lower left to the
upper right. Note that for a fair comparison between platforms, all gate times
have been weighted by each platform's intrinsic error time scales as described
previously. Circuits with run times significantly exceeding the platform's
intrinsic, level-dependent error time scales, highlighted by the gray area in
Fig.~\ref{fig:comp} (a) and (c), will most likely yield unreliable results. We
observe that despite the different time scales for gates, the weighted circuit
run times are almost identical for both platforms and for both the QFT and QAOA,
see squares in Fig.~\ref{fig:comp} (a) and (c), respectively. However, only
problem instances with relatively small qubit numbers seem to be currently
doable on both platforms.

While the circuit run time is one deciding factor in whether it is feasible on
current NISQ hardware, this measure neglects the fact that every gate comes with
an intrinsic error probability. Assuming gate errors on the order of $0.1-1\%$,
which are realistic both for neutral atoms~\cite{PRL.123.170503, Graham2022} and
superconducting circuits~\cite{datasheet}, it is clear that also the gate count
limits the feasibility of quantum circuits and lower gate counts are thus
preferable. To this end, the squares in Fig.~\ref{fig:comp} (b) and (d) show the
number of two-qubit gates required for QFT and single-step QAOA circuits,
respectively. Note that circuit representations in terms of the SGS require the
same number of two-qubit gates both on neutral atom and superconducting qubit
hardware. Hence, both cases are represented by a single line in
Figs.~\ref{fig:comp} (b) and (d).

\subsubsection{QSL gate set circuits}
The circles in Fig.~\ref{fig:comp} (a) and (c) show the circuit run times for
the same quantum algorithms and complexity levels as used for the squares but
when the QGS is employed to generate circuit representations. The corresponding
two-qubit gate counts are illustrated by the circles in Fig.~\ref{fig:comp} (b)
and (d). We observe an overall reduction in circuit run times and gate counts
for both platforms, both algorithms and any considered problem instance. This
improvement is a combined effect of using maximally fast quantum gates at the
QSL and taking advantage of the increased flexibility provided by the extended
set of gates. Especially the availability of SWAP gates needs to be stressed as
it directly reduces the gate count and thus leads to a reduction in circuit run
time even without an additional speedup in gate times. In order to quantify the
improvements achievable through the QGS, Fig.~\ref{fig:stats} (a) lists the
average reduction in circuit run times and gate counts for each algorithm and
platform. 
\par
Our analysis demonstrates to which extent circuit run times and gate counts can
--- from a theoretical perspective --- still be improved if standard gates are
replaced by an extended gate set with gate times at the QSL. However, even with
the improved gate set and its reduction in run time and gate count, most of the
quantum circuits, i.e., circles in Fig.~\ref{fig:comp}, remain likely infeasible
using current NISQ hardware. Quantum error correction codes could in principle
address the issue of circuit run times and gate counts exceeding the limits set
by finite lifetimes and gate errors. Nevertheless, since both
lifetimes~\cite{Barnes2022, Wang2022} and qubit numbers~\cite{ibmroadmap} are
constantly increasing, more complex quantum circuits will likely reach the
feasible regime in the near future.

\subsection{Circuits times in the parity mapping}
\label{subsec:parity}
Besides the representation of quantum algorithms using the SGM, we now consider
the representation of the same algorithms within the PM. While the PM was
originally designed to tackle combinatorial optimization problems via quantum
annealing~\cite{Lechner2015}, it can also be utilized for digital quantum
optimization algorithms such as QAOA~\cite{Ender2022} as well as to achieve
universal quantum computing~\cite{PRL.129.180503}. At its core, the PM
circumvents the need for long-range interactions between qubits, which in turn
renders gates between non-adjacent qubits obsolete. However, this comes at the
expense of requiring more physical qubits and many-body constraints on $2 \times
2$ plaquettes of qubits as specified in detail in
Appendix~\ref{app:circs:parity}. Similar to our analysis for the SGM (see
Sec.~\ref{subsec:standard}) we use either gates from the SGS or the QGS in the
following.

\subsubsection{Standard gate set circuits}
The plus signs in Fig.~\ref{fig:comp} (a) [(b)] show the circuit run times
[two-qubit gate counts] of QFTs in the PM for the same problem instances as used
for the SGM (squares). In both cases and for both platforms, the gates and
times of the SGS have been used. We observe a reduction in circuit run times and
gate counts for both platforms with the comparison being made with respect to
the values for the SGM [see also Fig.~\ref{fig:stats} (c)]. The same comparison
can be done for the QAOA steps with their circuit run times and gate counts
given by the plus signs in Fig.~\ref{fig:comp} (c) and (d), respectively.
Regarding single-step QAOA resource requirements, we observe a significant
reduction in circuit run times [see also Fig.~\ref{fig:stats} (c)] due to the
constant circuit depth in the PM~\cite{Lechner2020, Ender2022}. However, note
that the circuit depth differs for neutral atoms and superconducting circuits
due to the different coupling mechanisms between qubits with neutral atoms
having the deeper circuits, cf.\ Appendix~\ref{app:circs:parity}. The reduction
in circuit run time is accompanied by an increase in two-qubit gate counts
compared to the SGM, which we attribute to the decomposition of the constraint
gates, cf.\ Eq.~\eqref{eq:Zk}, into the natively available gates within each
platform. 

\subsubsection{QSL gate set circuits}
The crosses in Fig.~\ref{fig:comp} show the results for circuit representations
in the PM when employing gates and times from the QGS. For the QFT on neutral
atoms, we do not observe any further reduction in circuit time or gate counts
compared to its representation utilizing the SGS. This is because its circuit
representations~\cite{PRA.106.042442} for neutral atoms contain only
single-qubit and CZ gates --- gates for which the gate times are identical
within the SGS and the QGS. In contrast, for superconducting circuits, we still
observe an improvement in circuit time since the CZ gate becomes directly
available in the QGS and must no longer be replaced by Sycamore and single-qubit
gates. However, the representations in the SGS and QGS only differ by
single-qubit gates and hence their two-qubit gate count is identical and also
identical to that of the neutral atoms. This is reflected in
Fig.~\ref{fig:comp} (c) by a single line of superimposed plus signs and crosses.
\par
For the single-step QAOA circuits in the PM, we observe a significant reduction
both in circuit run time and gate counts on both platforms when the QGS is used
instead of the SGS. This is due to the availability of the multi-qubit gates
$\mathrm{ZZZ}(\gamma)$ and $\mathrm{ZZZZ}(\gamma)$, cf.~Eq.~\eqref{eq:Zk}, where
the gate-count reduction originates from avoiding single- and two-body gate
decompositions. In addition, it turns out to be much faster to use control
pulses that directly implement these multi-qubit gates as opposed to serially
applying control pulses to implement the required single- and two-qubit gates.
Figure~\ref{fig:stats} (b) summarizes the average run time and gate count
reductions (in terms of two- and multi-qubit gates) when replacing the SGS with
the QGS within the PM.
\par
In contrast to panels (a) and (b) of Fig.~\ref{fig:stats}, where the gate set
changes and the circuits stay in the SGM or PM, panels (c) and (d) examine the
opposite scenario, i.e., the gate sets are kept constant but the circuit models
change. In detail, Fig.~\ref{fig:stats} (c) and (d) show the average circuit run
time and gate count reduction when the SGM is replaced by the PM while the gate
set is given by the SGS and QGS, respectively. Especially Fig.~\ref{fig:stats}
(d) needs to be emphasized as it reveals to which extent the PM allows to reduce
circuit run times and gate counts compared to representations of the same
circuits in the SGM. As mentioned above and detailed in
Appendix~\ref{app:circs:parity}, the PM-specific improvements come at the only
expense of requiring more qubits. For the current state of NISQ hardware and
independent from the platform, we believe the SGM to be better suited for QFTs,
as more problem instances seem to be feasible, judging by circuit run time and
required qubits. In contrast, for QAOA using the QGS, we find the described
parity representation of circuits (despite its overhead-induced inferior success
probabilities compared to direct SGM-QAOA~\footnote{Note that this statement
does not include additional PM-QAOA-performance-increasing strategies such as
described in Ref.~\cite{Weidinger2023}}) to be an advantageous option, as the
run time and gate count is drastically reduced for each QAOA step compared to
the SGM. In particular, the need for more PM-QAOA steps compared to SGM-QAOA
steps might be compensatable given the resource reduction per PM-QAOA step.
Given the ongoing upscaling of quantum computers in terms of qubit numbers, see,
e.g., IBM's quantum roadmap~\cite{ibmroadmap}, the PM seems to be a viable
option for QAOA.

As a final remark, it should be noted that we only discussed optimization
problems with all-to-all connectivity, cf.\ Eq.~\eqref{eq:Hz}. However, many
realistic optimization problems have rather sparse connectivity, which would
lead to a significant reduction in required qubits~\cite{Ender2023} while
maintaining its strength of constant circuit depth.

\section{Modeling neutral atom and superconducting circuit platforms}
\label{sec:model}
In Sec.~\ref{sec:main} we have presented the main result of our work --- namely
a calculation and comparison of circuit run times and gate counts using neutral
atoms and superconducting circuits as quantum computing platform. In this
section, we now introduce the detailed physical models used for both platforms.
Since our focus is the description of dynamics, we focus on the respective
Hamiltonians including the various control knobs typically available to steer
the systems and implement quantum gates.

\subsection{Neutral atoms}
\label{subsec:ryd_model}
Arrays of trapped neutral atoms laser coupled to highly excited Rydberg states
are a promising platform for quantum computing~\cite{Saffman2016, Henriet2020}
and quantum simulation~\cite{Browaeys2020} as qubits can for example be encoded
in long-lived hyperfine ground states. High-fidelity single-qubit gates can be
achieved using microwave fields~\cite{Schrader2004}, two-photon Raman
transitions~\cite{Yavuz2006}, or a combination of microwaves and gradient fields
for individual-qubit addressing~\cite{PRL.114.100503, PRL.115.043003, Wang2016}.
In contrast, entangling operations between atoms, i.e., many-body gates, are
typically realized via strongly interacting Rydberg levels~\cite{Jaksch2000,
PRL.87.037901} and various control schemes for two- and multi-qubit gates have
been experimentally demonstrated~\cite{PRA.92.022336, Jau2015, PRL.121.123603,
PRL.123.170503, PRL.123.230501}.
\par
Such gates have been used in recent experiments~\cite{Bluvstein2022, Graham2022}
with up to several hundreds of atoms arranged in a planar geometry. If we
consider the smallest building block of such a 2D array, it consists of $N=4$
atoms with each atom described by three relevant levels. In the rotating frame,
its Hamiltonian reads ($\hbar = 1$)
\begin{equation} \label{eq:ham_ryd}
  \begin{split}
      \op{H}(t)
    &=
    - \sum_{n=1}^{N} \Delta_{n}(t) \ket{r_{n}} \bra{r_{n}}
    +
    \sum_{\substack{n,m=1 \\ n < m}}^{N}
    V_{nm} \ket{r_{n} r_{m}} \bra{r_{n} r_{m}}
     \\ &\quad
    +
    \frac{1}{2} \sum_{n=1}^{N} \sum_{l=\uparrow,\downarrow} \Big[
      \Omega_{l,n}(t) e^{\im \varphi_{l,n}} \ket{r_{n}} \bra{l_{n}}
      +
      \mathrm{H.c.}
    \Big],
  \end{split}
\end{equation}
where $\ket{\downarrow_{n}}$ and $\ket{\uparrow_{n}}$ denote the qubit levels of
the $n$th atom and $\ket{r_{n}}$ its Rydberg level. The Rabi frequencies of the
two laser fields coupling the $\ket{\downarrow_{n}}$ and $\ket{\uparrow_{n}}$
states of the $n$th atom to their respective Rydberg level $\ket{r_{n}}$ are
denoted by $\Omdn(t)$ and $\Omun(t)$~\footnote{In practice, the coupling between
qubit states and the Rydberg level is typically achieved via a two-photon
process and an intermediate level~\cite{Bernien2017}, which we neglect for
simplicity}. The phases of the respective laser fields are given by $\phidn(t)$
and $\phiun(t)$ while $\Delta_{n}(t)$ denotes the detuning of the laser field
from the exact transition frequency. We assume $\Delta_{n}(t)$ to be identical
for both fields. The van der Waals (vdW) interaction between atom $n$ and $m$ is
denoted by $V_{nm}$. Note that in Hamiltonian~\eqref{eq:ham_ryd} we have omitted
fields responsible for single-qubit gates, i.e., fields coupling the two qubit
levels, since our focus is on multi-qubit gates for which the Rydberg levels and
their interaction are the primarily gate mechanism.

If not stated otherwise, we assume the laser fields to act identically on all
atoms, i.e., $\Omdn(t) = \Omd(t)$, $\Omun(t) = \Omu(t)$ and $\phidn(t)
= \phid(t)$, $\phiun(t) = \phiu(t)$ and $\Delta_{n}(t) = \Delta(t)$ for all $n$.
In the following, we moreover assume by default a pseudo-2D architecture with
$V_{nm} = V$ as described in more detail in Ref.~\cite{PRL.128.120503}.
Effectively, this corresponds to an architecture with equally strong nearest
neighbor (NN) and next-to-nearest neighbor (NNN) coupling. However, it should be
noted that we do not exploit the NNN couplings when constructing the quantum
circuits in Sec.~\ref{sec:main}. It thus does not affect any of the single- or
two-qubit gate times presented in Table~\ref{tab:gatesets} as these times would
be identical for an actual, planar 2D architecture where $V$ is simply the
coupling strength between nearest neighbors. The assumption of of the pseudo-2D
architecture does only affect the multi-qubit constraint gates
$\mathrm{ZZZ}(\gamma)$ and $\mathrm{ZZZZ}(\gamma)$, cf.\ Eq.~\eqref{eq:Zk},
which are --- within our study --- only relevant for the PM using the QGS, cf.\
the crosses in Fig.~\ref{fig:comp} (c).

\subsection{Superconducting Circuits}
\label{subsec:sc_model}
Qubits encoded in the lowest energy levels of superconducting circuits are
another promising platform for quantum computing~\cite{Kjaergaard2020}. Since
their parameters can be to some extent chosen during their fabrication process,
superconducting circuits come in various variants and parameter regimes with
transmon qubits~\cite{PRA.76.042319} being currently the most prominent ones for
quantum computing. Their qubit levels can be manipulated via microwave fields to
either implement single-qubit gates~\cite{PRL.110.040502, PRApplied.7.041001} or
two-qubit gates~\cite{Barends2014, PRX.11.021058} with high fidelity. In
contrast to neutral atom qubits, which interact directly via their Rydberg
levels, qubits encoded in superconducting circuits often interact indirectly via
intermediate coupling elements~\cite{Majer2007}. In an architecture, where such
couplers are made tunable~\cite{PRL.113.220502}, the effective interaction
strength between the qubits becomes tunable as well. Such qubit architectures
have been successfully used in recent quantum advantage
experiments~\cite{Arute2019, PRL.127.180501} and thus are a prototypical NISQ
quantum computing platform.

The qubits in such a tunable coupler architecture are arranged on a 2D lattice
with nearest-neighbor couplings. In the following, we take the architecture from
Ref.~\cite{Arute2019} as a reference. The smallest building block within such
a system consists of $N=4$ qubits. The Hamiltonian for this subsystem
reads~\cite{Arute2019}
\begin{equation} \label{eq:ham_sc}
  \begin{split}
    \op{H}(t)
    &=
    \sum_{n=1}^{N} \left[%
      \omega_{n}(t) \op{b}_{n}^{\dagger} \op{b}_{n}
      -
      \frac{\alpha_{n}}{2}
      \op{b}_{n}^{\dagger} \op{b}_{n}^{\dagger} \op{b}_{n} \op{b}_{n}
    \right]
    \\
    &\quad
    +
    \sum_{n=1}^{N}
    \left(\op{b}_{n} + \op{b}_{n}^{\dagger}\right)
    \Omega_{n}(t) \cos(\bar{\omega}_{n}(t) t)
    +
    \op{H}_{\rmint}(t)
  \end{split}
\end{equation}
with
\begin{equation} \label{eq:ham_sc_int}
  \begin{split}
    \op{H}_{\rmint}(t)
    &=
    g_{12}(t)
    \left(\op{b}_{1}^{\dagger} \op{b}_{2} + \op{b}_{1} \op{b}_{2}^{\dagger}\right)
    +
    \frac{g_{12}^{2}(t)}{|\eta|}
    \op{b}_{1}^{\dagger} \op{b}_{1} \op{b}_{2}^{\dagger} \op{b}_{2}
    \\
    &\quad
    +
    g_{23}(t)
    \left(\op{b}_{2}^{\dagger} \op{b}_{3} + \op{b}_{2} \op{b}_{3}^{\dagger}\right)
    +
    \frac{g_{23}^{2}(t)}{|\eta|}
    \op{b}_{2}^{\dagger} \op{b}_{2} \op{b}_{3}^{\dagger} \op{b}_{3}
    \\
    &\quad
    +
    g_{34}(t)
    \left(\op{b}_{3}^{\dagger} \op{b}_{4} + \op{b}_{3} \op{b}_{4}^{\dagger}\right)
    +
    \frac{g_{34}^{2}(t)}{|\eta|}
    \op{b}_{3}^{\dagger} \op{b}_{3} \op{b}_{4}^{\dagger} \op{b}_{4}
    \\
    &\quad
    +
    g_{41}(t)
    \left(\op{b}_{4}^{\dagger} \op{b}_{1} + \op{b}_{4} \op{b}_{1}^{\dagger}\right)
    +
    \frac{g_{41}^{2}(t)}{|\eta|}
    \op{b}_{4}^{\dagger} \op{b}_{4} \op{b}_{1}^{\dagger} \op{b}_{1},
  \end{split}
\end{equation}
where $\omega_{n}(t)$ is the frequency-tunable level splitting of transmon $n$,
$\alpha_{n}$ its anharmonicity and $\op{b}_{n}$ its annihilation operator. The
tunable coupling between transmons $n$ and $m$ is denoted by $g_{nm}(t)$ and
$\eta$ is the non-linearity of the involved transmons, which is roughly constant
$\alpha_{n} \approx \eta$ for all $n$. The time-dependent amplitude and
frequency of a local $X$-type control field on transmon $n$, e.g., some
microwave field, is denoted by $\Omega_{n}(t)$ and $\bar{\omega}_{n}(t)$,
respectively. For numerical reasons, it is advantageous to change into
a rotating frame. We change into a rotating frame with frequency
$\omega_{\rmrot}$ via the transformation
\begin{subequations} \label{eq:sc_trans}
  \begin{align}
    \op{H}'(t)
    &=
    \op{O}^{\dagger}(t) \op{H}(t) \op{O}(t)
    - \im \op{O}^{\dagger}(t) \frac{\dd \op{O}(t)}{\dd t},
    \\
    \op{O}(t)
    &=
    \exp\left\{%
      - \im \omega_{\rmrot} \left(%
        \sum_{n=1}^{N} \op{b}_{n}^{\dagger} \op{b}_{n}
      \right) t
    \right\}
  \end{align}
\end{subequations}
and find
\begin{equation} \label{eq:ham_sc'}
  \begin{split}
    \op{H}'(t)
    &=
    \sum_{n=1}^{N} \left[%
      \left(\omega_{n}(t) - \omega_{\rmrot}\right) \op{b}_{n}^{\dagger} \op{b}_{n}
      -
      \frac{\alpha_{n}}{2}
      \op{b}_{n}^{\dagger} \op{b}_{n}^{\dagger} \op{b}_{n} \op{b}_{n}
    \right]
    \\
    &\quad+
    \sum_{n=1}^{N} \frac{1}{2} \left[%
      \op{b}_{n}
      \left(\bar{\Omega}_{n,\rmre}(t) + \im \bar{\Omega}_{n,\rmim}(t)\right)
      \right. \\ &\qquad \qquad \left.
      +
      \op{b}_{n}^{\dagger}
      \left(\bar{\Omega}_{n,\rmre}(t) - \im \bar{\Omega}_{n,\rmim}(t)\right)
    \right]
    +
    \op{H}_{\rmint}(t),
  \end{split}
\end{equation}
where we have introduced the auxiliary control fields
\begin{subequations}
  \begin{align}
    \bar{\Omega}_{n,\rmre}(t)
    &=
    \rp\left\{%
      \Omega_{n}(t) e^{- \im (\omega_{\rmrot} - \bar{\omega}_{n}(t)) t}
    \right\},
    \\
    \bar{\Omega}_{n,\rmim}(t)
    &=
    \ip\left\{%
      \Omega_{n}(t) e^{- \im (\omega_{\rmrot} - \bar{\omega}_{n}(t)) t}
    \right\}.
  \end{align}
\end{subequations}
While $\Omega_{n}(t)$ and $\bar{\omega}_{n}(t)$ are the actual physical control
fields, we may take $\bar{\Omega}_{n,\rmre}(t)$ and $\bar{\Omega}_{n,\rmim}(t)$
as auxiliary control fields which capture the time-dependent nature of
$\Omega_{n}(t)$ and $\bar{\omega}_{n}(t)$ in the rotating frame and the latter
can be reobtained from $\bar{\Omega}_{n,\rmre}(t)$ and
$\bar{\Omega}_{n,\rmim}(t)$.

\section{Determining quantum speed limits via quantum optimal control}
\label{sec:oct}
After having introduced the physical model for neutral atoms and superconducting
circuits in Sec.~\ref{sec:model}, we now review the basic notion of QSLs in
Sec.~\ref{subsec:qsl} and of quantum optimal control theory in
Sec.~\ref{subsec:oct} since both build the theoretical, respectively methodical,
foundation for the results presented in this work. Our method is described in
Sec.~\ref{subsec:method}.

\subsection{Quantum speed limits}
\label{subsec:qsl}
The notion of quantum speed limits (QSLs) naturally arises in the context of
quantum control problems. To this end, let us consider a quantum system
described by the Hamiltonian $\op{H}(t) = \op{H}(\{\pulse_{k}(t)\})$, which
depends on a set of control fields, $\{\pulse_{k}(t)\}$, that can be externally
tuned, e.g., by the time-dependent amplitudes, phases or detunings in
Eqs.~\eqref{eq:ham_ryd} or \eqref{eq:ham_sc'}. A quantum control problem is then
defined by a set of initial states, $\{\ket{\psi_{l}^{\rmin}}\}$, that should be
transferred into a set of target states, $\{\ket{\psi_{l}^{\trgt}}\}$,
\begin{align} \label{eq:trans}
  \Ket{\psi_{l}^{\trgt}}
  =
  \op{U}(T,0; \{\pulse_{k}(t)\}) \Ket{\psi_{l}^{\rmin}},
  \quad \forall l,
\end{align}
where $\op{U}(T,0; \{\pulse_{k}(t)\})$ is the system's time-evolution operator
and $T$ the total time. Any choice of $\{\pulse_{k}(t)\}$ which fulfills
Eq.~\eqref{eq:trans} is considered a solution to the control problem. It is
important to note that solutions to quantum control problems are usually not
unique. Even for a fixed protocol duration $T$ there typically exist many and
often infinite many solutions.

The QSL for a given control problem is defined by the shortest protocol duration
$T_{\rmqsl}$ for which at least one solution exists, i.e., for which at least
one set of control fields $\{\pulse_{k}(t)\}$ exists that fulfills
Eq.~\eqref{eq:trans}. In the context of quantum computing and the NISQ era,
where time is a limited resource due to decoherence, it is desirable to
implement quantum gates at the QSL. In order to calculate $T_{\rmqsl}$
analytically, $\op{U}(T,0; \{\pulse_{k}(t)\})$ must be analytically calculable
for \emph{any} set $\{\pulse_{k}(t)\}$ of conceivable control fields ---
a requirement that typically limits an analytical calculation of $T_{\rmqsl}$ to
simple systems~\cite{PRXQuantum.2.030203}.

Besides an analytical determination, there are various methods to approximate
$T_{\rmqsl}$. One prominent method is to calculate a lower bound $T_{\rmbound}
\leq T_{\rmqsl}$ in order to get an estimate for $T_{\rmqsl}$ itself. For the
simplest case of a state-to-state control problem, lower bounds can be
calculated analytically~\cite{Deffner2017}. In contrast, in case of multiple
pairs of initial and final states, which describe the implementation of quantum
gates, such lower bounds only exist for very simple systems~\cite{Svozil2005}.
In most cases, one needs to resort to numerical tools for estimating
$T_{\rmqsl}$. In that context, quantum optimal control theory has proven to be
very useful~\cite{PRL.103.240501} as it not only estimates $T_{\rmqsl}$ quite
accurately but additionally yields the control fields that implement the desired
dynamics, i.e., realizing the transition from initial to target states, cf.\
Eq.~\eqref{eq:trans}. Since this is our method of choice, we introduce it in
more detail in the following.

\subsection{Quantum optimal control theory}
\label{subsec:oct}
Quantum optimal control theory (OCT)~\cite{Koch2022} is a toolbox providing
analytical and numerical tools that allow to derive optimized control fields
which solve a given control problem, e.g., in shortest time or with minimal
error. Mathematically, an optimal control problem is formulated by introducing
the cost functional
\begin{equation} \label{eq:J}
  \begin{split}
      J\left[\{\psi_{l}\}, \{\pulse_{k}\}, T\right]
      &=
      \error_{T}\left[\{\psi_{l}(T)\}\right]
      \\ &\quad
      +
      \int_{0}^{T} J_{t}\left[\{\psi_{l}(t)\}, \{\pulse_{k}(t)\}, t\right] \dd t,
  \end{split}
\end{equation}
where $\{\psi_{l}(t)\}$ is a set of time-evolved states and $\{\pulse_{k}(t)\}$
a set of control fields to be optimized. The error-measure $\error_{T}$
quantifies the distance between the time-evolved states $\ket{\psi_{l}(T)}
= \op{U}(T,0; \{\pulse_{k}(t)\}) \ket{\psi_{l}^{\rmin}}$ and the desired target
states $\ket{\psi_{l}^{\trgt}}$ at the protocol's final time $T$, cf.\
Eq.~\eqref{eq:trans}. The term $J_{t}$ in Eq.~\eqref{eq:J} captures
time-dependent running costs. In most cases, the error-measure $\error_{T}$ is
the crucial figure of merit. In order to optimize for quantum gates, we use the
error-measure~\cite{PRA.68.062308}
\begin{align} \label{eq:error}
  \error_{T}\left[\{\psi_{l}(T)\}\right]
  =
  1 - \frac{1}{N_{\trgt}} \sum_{l=1}^{N_{\trgt}}
  \rp\left\{%
    \Braket{\psi_{l}^{\trgt} | \psi_{l}(T)}
  \right\}
\end{align}
and take $\ket{\psi_{l}^{\trgt}} = \op{O} \ket{\psi_{l}^{\rmin}}$ as the desired
target states for the target gate $\op{O}$. The set $\{\psi_{l}^{\rmin}\}$ runs
over the $N_{\trgt}$ logical basis states affected by $\op{O}$ with $N_{\trgt}
= 4, 8, 16$ for two-, three- or four-body gates, respectively.

Since the cost functional $J$ is formulated such that smaller values correspond
to better solutions of the control problem, solving an optimal control problem
becomes essentially a minimization task, i.e., to find a set of control fields
$\{\pulse_{k}^{\rmopt}(t)\}$ that minimizes $J$, respectively $\error_{T}$. This
is an optimization problem for which several numerical algorithms have been
developed~\cite{JMagRes.172.296, CanevaPRA2011, JCP.136.104103, SorensenPRA2018,
MachnesPRL2018}. Many of them are readily available in open-source software
packages~\cite{SciPostPhys.7.6.080, PRApplied.15.034080, Goerz2022,
PRApplied.17.034036, Rossignolo2022}.

\subsection{Optimization procedure and method}
\label{subsec:method}
In the context of determining $T_{\rmqsl}$ numerically via OCT, we search for
the shortest protocol duration $T$ for which the error-measure $\error_{T}$,
cf.\ Eq.~\eqref{eq:error}, is still sufficiently small. In mathematical terms,
we therefore define an error threshold $\error_{\rmmax}$ and search for the
shortest time $T$ for which
\begin{align} \label{eq:qsl}
  \underset{\{\pulse_{k}\}}{\rmmin}
  \left[
    \error_{T}\left[\{\psi_{l}(T)\}\right]
  \right]
  \leq
  \error_{\rmmax}
\end{align}
has a solution. However, the minimization over all conceivable control fields
can not be done numerically --- as there are infinitely many fields to check ---
and thus must be replaced by a sampling over finitely many fields in practice.
In order to explore the function space efficiently by finite sampling,
optimization algorithm as described in Sec.~\ref{subsec:oct} can be used. Since
we are interested in the fundamental QSL, we put only minimal limitations ---
apart from physically motivated limitations on amplitudes --- on the form of
each control field $\pulse_{k}(t)$. Hence, we need an optimization algorithm
that is capable of exploring a function space of almost arbitrary field shapes.

As our method of choice we use Krotov's method~\cite{AutomRemContr.60.1427},
a gradient-based optimization algorithms for time-continuous control fields.
While a more detailed description of Krotov's method is given in
Appendix~\ref{app:krotov}, its basic working principle is outlined in the
following. It consists of an iterative update of the control fields
$\{\pulse_{k}(t)\}$. Starting from a set of guess control fields
$\{\pulse_{k}^{0}(t)\}$, Krotov's method updates them until either $\error_{T}
\leq \error_{\rmmax}$ or a maximum number of iterations is reached. This
procedure can be viewed as a local but structured search within the space of all
conceivable sets of control fields --- the so-called control landscape. The
locally searched area is thereby determined by the choice of the guess fields
$\{\pulse_{k}^{0}(t)\}$, which set the initial starting point of the search.
While the local nature of this search might appear contradicting to the global
search required for evaluating Eq.~\eqref{eq:qsl}, it can be turned into an
approximate global search by using various sets of randomized guess fields. The
combined effect of all these local searches ``cover'' a larger fraction of the
control landscape. The total procedure for approximating $T_{\rmqsl}$ using this
method is thus to start with a protocol duration $T$ for which the optimization
algorithm finds solutions, i.e., optimized fields giving rise to $\error_{T}
\leq \error_{\rmmax}$, and then to consecutively lower $T$ until none of the
various sets of randomized guess fields finds a solution anymore.
Appendix~\ref{app:field} summarizes the details regarding the generation of
random guess fields as well as each field's parametrization within Krotov's
method.

Similar application of numerical optimal control techniques have previously
shown excellent agreement with analytically provable QSLs~\cite{PRA.92.062110,
PRR.3.013110}. At worst, this method could overestimate the actual QSLs, in
which case the actual QSLs would be even smaller and the circuit and gate times
in Sec.~\ref{sec:main} that use the QGS would be even better.


\begin{figure*}[!tb]
  \centering
  \includegraphics{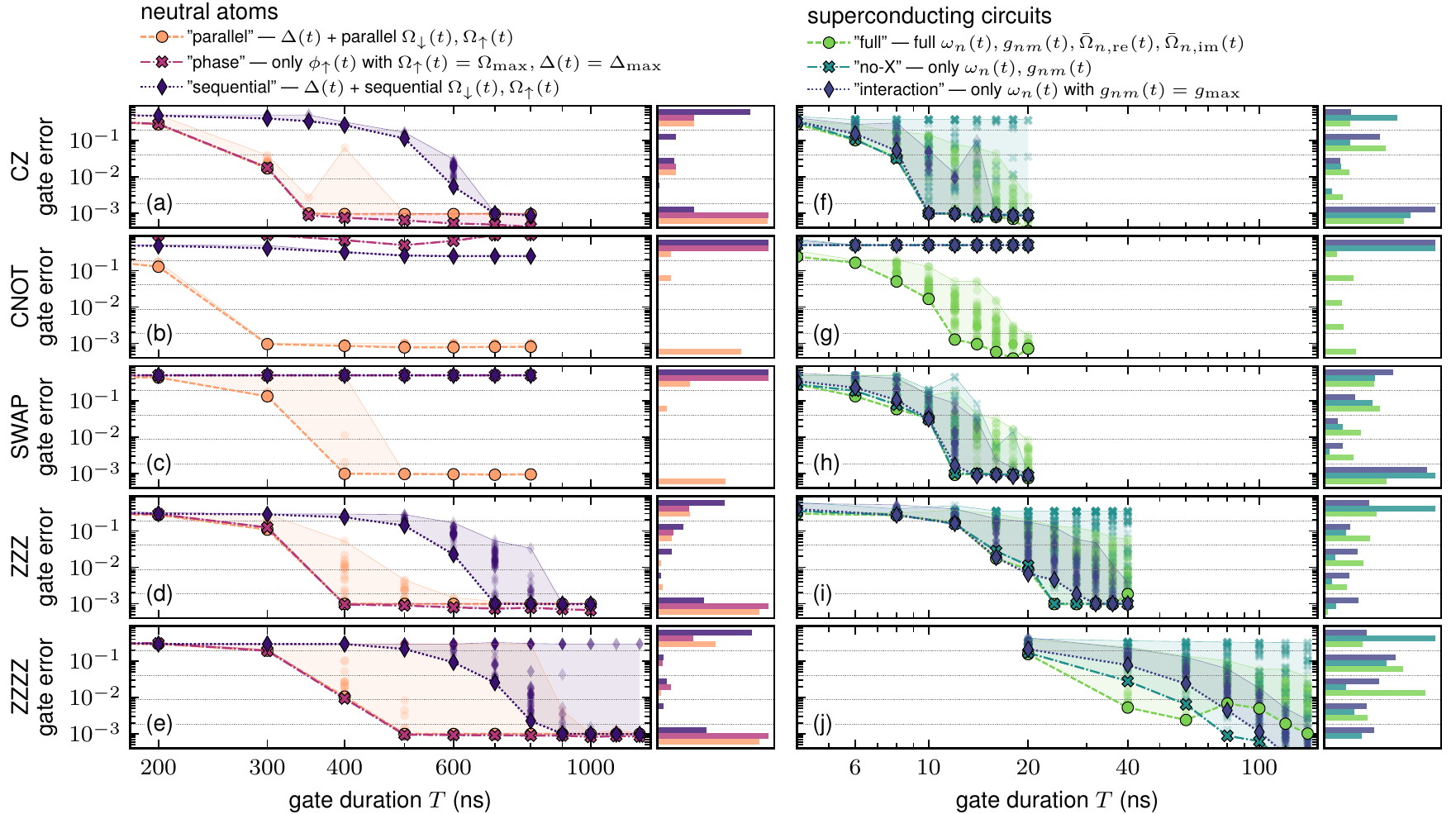}
  \caption{%
    Overview of the QSLs for the various gates of the ``QSL gate set'' (QGS) in
    Table~\ref{tab:gatesets}. The left (right) column shows the results for
    neutral atoms (superconducting circuits) for three different configurations
    of control fields (specified in the main text), respectively. Each marker
    represents the gate error $\error_{T}$, cf.\ Eq.~\eqref{eq:error}, after
    either reaching $\error_{T} \leq \error_{\rmmax} = 10^{-3}$ or $1500$
    iterations of Krotov's method, cf.\ Appendix~\ref{app:krotov}, after
    starting from a set of random guess fields generated via Eq.~\eqref{eq:f}.
    While the lines connect the lowest errors reached for each gate time $T$
    within a given field configuration, the shaded background color indicates
    the range between the lowest and highest error. The marker density is shown
    at the right side of each panel as a histogram. The parameters for neutral
    atoms are $V/2\pi = \SI{40}{\mega\hertz}$, $0 \leq \Omd(t), \Omu(t) \leq
    \Omega_{\rmmax} = 0.1 V$ and $|\Delta(t)| \leq \Delta_{\rmmax} = 0.3 V$. The
    parameters for superconducting circuits are $-\SI{40}{\mega\hertz} \leq
    g_{nm}(t)/2\pi \leq \SI{5}{\mega\hertz}$, $\SI{6700}{\mega\hertz} \lesssim
    \omega_{n}(t) \lesssim \SI{7100}{\mega\hertz}$ (exact values depending on
    $n$ and taken from Ref.~\cite{Arute2019}) and $-\SI{50}{\mega\hertz} \leq
    \bar{\Omega}_{n, \rmre}(t), \bar{\Omega}_{n, \rmim}(t) \leq
    \SI{50}{\mega\hertz}$. The anharmonic ladder for each transmon has been
    truncated after five levels with population in the highest level suppressed
    during optimization. The optimization results for $\mathrm{ZZZ}(\gamma)$ and
    $\mathrm{ZZZZ}(\gamma)$, cf.\ Eq.~\eqref{eq:Zk}, have been obtained for the
    maximally entangling gate at $\gamma = \pi/4$. The results for the CNOT gate
    on neutral atoms have been obtained using site-dependent control fields for
    each atom.
  }
  \label{fig:qsl}
\end{figure*}

\section{Benchmarking quantum gate times on 2D architectures}
\label{sec:bench}
In this final section, we now present the detailed results regarding the QSLs
obtained via methods described in Sec.~\ref{sec:oct} for the various gates
listed in Table~\ref{tab:gatesets} and have been used to calculate the circuit
run times in Sec.~\ref{sec:main}.

\subsection{Neutral atoms}
\label{subsec:ryd}
In this section, we determine the QSLs for different gates on neutral atoms by
utilizing the control knobs available in Hamiltonian~\eqref{eq:ham_ryd}. To this
end, we set the vdW interaction strength between Rydberg levels to $V/2\pi
= \SI{40}{\mega\hertz}$ and assume a maximally achievable Rabi frequency of
$\Omega_{\rmmax}/2\pi = 0.1 V = \SI{4}{\mega\hertz}$ for both $\Omd(t)$ and
$\Omu(t)$ as well as $\Delta_{\rmmax}/2\pi = 0.3 V = \SI{12}{\mega\hertz}$ for
$|\Delta(t)|$. These parameters are in the same regime than those reported in
recent experiments~\cite{Bernien2017, PRL.123.170503, Bluvstein2022}.

The markers in Fig.~\ref{fig:qsl} (left column) show the achievable gate error
$\error_{T}$, cf.\ Eq.~\eqref{eq:error}, for various gates and various gate
times $T$ on neutral atoms. Each individual marker thereby indicates the result
of a single optimization with Krotov's method [cf.\ Appendix~\eqref{app:krotov}]
i.e., the final error $\error_{T}$ after $1500$ iterations when starting from
random guess fields generated via Eq.~\eqref{eq:f}. In the following we set
$\error_{\rmmax} = 10^{-3}$ for all gates and stop any optimization as soon as
this threshold is reached.

\subsubsection{Two-qubit gates}
Figure~\ref{fig:qsl} (a) shows the results for a CZ gate for three different,
paradigmatic configurations of control fields. The circles correspond to the
``parallel'' configuration where all five possible control fields $\Omd(t),
\Omu(t), \phid(t), \phiu(t)$ and $\Delta(t)$ have been optimized. In contrast,
the crosses correspond to the ``phase'' configuration, where only $\phiu(t)$ has
been optimized while $\Omu(t) = \Omega_{\rmmax}$, $\Delta(t) = \Delta_{\rmmax}$
and $\Omd(t) = \phid(t) = 0$ have been kept fixed. In both cases, we obtain
a QSL of $T_{\rmqsl}^{\mathrm{CZ}} = \SI{350}{\nano\second}$. The third field
configuration, which we called ``sequential'' configuration (diamonds), consists
of a sequential use of $\Omd(t), \phid(t)$ and $\Omu(t), \phiu(t)$, i.e., the
first half of the protocol we have $\Omu(t) = \phiu(t) = 0$ and in the second
half $\Omd(t) = \phid(t) = 0$. This configuration is inspired by an adiabatic
protocol for implementing $\mathrm{ZZZZ}(\gamma)$ gates, cf.\ Eq.~\eqref{eq:Zk}
and Ref.~\cite{PRL.128.120503}. Its QSL $T_{\rmqsl}^{\mathrm{CZ}}
= \SI{700}{\nano\second}$ is twice as long compared to the other two
configurations.

In order to compare the results from the three configurations in terms of how
successful the optimization has been in finding solutions, the histogram on the
right side of Fig.~\ref{fig:qsl} (a) provides the probability density for
obtaining final errors $\error_{T}$ within certain ranges. For obtaining
a solution with errors $\error_{T} \leq \error_{\rmmax}$, we find the lowest
probability for the ``sequential'' configuration (diamonds) --- coinciding with
the highest QSL --- and the highest and almost identical probability for the
other two configurations. Among those two, the ``phase'' configuration (crosses)
needs to be emphasized in particular. From a physical perspective, setting
$\Omd(t)$ and $\phid(t)$ to zero automatically ensures that $\ket{\downarrow
\downarrow}$ is mapped onto itself --- as required by the CZ gate. This is not
automatically guaranteed by the other two configurations and the optimization
needs to explicitly ensure it and therefore needs to solve a slightly more
complex optimization problem. However, the advantage of having one basis state
automatically mapped correctly does not translate into an advantage regarding
the QSL of $T_{\rmqsl}^{\mathrm{CZ}} = \SI{350}{\nano\second}$ or the reachable
error in general. Interestingly, both the ``parallel'' and ``phase''
configurations yield the same achievable lowest errors for each gate time $T$.
This is visually highlighted by the lines connecting the lowest errors per $T$
in Fig.~\ref{fig:qsl} (a). Moreover, it should be stressed that these errors
$\error_{T}$ are reached for almost every set of initial guess fields, i.e.,
independent of the initial starting point within the problem's control
landscape, and obtained independently for both configurations. This supports the
conjecture that $T_{\rmqsl}^{\mathrm{CZ}} = \SI{350}{\nano\second}$ is the
actual QSL for a CZ gate and generally validates our method in determining the
QSL. From an optimal control perspective, it is interesting to see that the
flexibility originating from the extended set of available control fields in the
``parallel'' configuration can not be turned into an advantage in error or time
compared to the ``phase'' configuration. From a practical perspective, the
latter is advantageous for experimental realizations as it requires fewer
physical resources.

While the three configurations discussed so far should only be viewed as
examples, we did not find any configuration giving rise to faster CZ gates.
Hence, we assume $T_{\rmqsl}^{\mathrm{CZ}} = \SI{350}{\nano\second}$ to be the
fundamental QSL across all configurations. A natural comparison for
$T_{\rmqsl}^{\mathrm{CZ}}$ with a value from the literature would be the gate
time from the analytical protocol introduced in Ref.~\cite{PRL.123.170503},
especially because it uses the same control fields as the ``phase''
configuration to implement the gate. For our parameters, we find
$T_{\rmlit}^{\mathrm{CZ}} \approx \SI{340}{\nano\second}$ as analytical gate
time, which we assume identical with our QSL $T_{\rmqsl}^{\mathrm{CZ}}
= \SI{350}{\nano\second}$ given the rather coarse sampling of gate times $T$ in
Fig.~\ref{fig:qsl} (a). However, it should be noted that the analytical protocol
of Ref.~\cite{PRL.123.170503} implements a CZ gate only up to local operations
--- operations that are already contained in our optimized gate protocols at the
QSL. Nevertheless, for a fair comparison of analytical and QSL gate times, as
needed in Sec.~\ref{sec:main}, as well as for simplicity, we set both times to
$\SI{350}{\nano\second}$ in Table~\ref{tab:gatesets}.


The remaining panels (b)-(e) in the left column of Fig.~\ref{fig:qsl} show the
results for other gates from the QGS of Table~\ref{tab:gatesets}. The results
for a CNOT gate are shown in panel (b). It is the only gate for neutral atoms
(among those we considered) that requires individual instead of global fields,
i.e., it is the only gate for which we did not assume $\Omdn(t) = \Omd(t),
\Omun(t) = \Omu(t), \phidn(t) = \phid(t), \phiun(t) = \phiu(t)$ and
$\Delta_{n}(t) = \Delta(t)$ for all $n$ but actually assume individual fields
with unique field shapes directed at each atom. We nevertheless consider the
same three field configurations as for the CZ gate in panel (a) but now applied
to the individual fields $\Omdn(t), \Omun(t), \phidn(t), \phiun(t)$ and
$\Delta_{n}(t)$ instead of their global versions. Even with this more general
setting of control fields, we find only the ``parallel'' configuration (circles)
to allow for the realization of a CNOT gate with a QSL of
$T_{\rmqsl}^{\mathrm{CNOT}} = \SI{300}{\nano\second}$. In contrast, the
``phase'' and ``sequential'' configurations do not allow to realize a CNOT gate
at all. These results demonstrate that CNOT gates can be implemented using
exclusively the site-dependent laser couplings of the qubit and Rydberg levels
and no control knobs for single-qubit gates. However, an experimentally more
convenient option, requiring no full site-dependent control of coupling qubit to
Rydberg states, is to realize CZ gates with global laser pulses and convert them
into CNOTs via local operations. We nevertheless include the CNOT gate and its
QSL for completeness in our analysis as well as in the QGS in
Table~\ref{tab:gatesets} but exclude it from any quantum circuit for neutral
atoms in Sec.~\ref{sec:main} for the reasons just mentioned.

Figure~\ref{fig:qsl} (c) shows the results for a SWAP gate. Like for the CNOT
gate, we find the ``parallel'' configuration --- assuming again global fields
that are identical for each atom --- to be the only one capable of realizing
a SWAP gate. We obtain $T_{\rmqsl}^{\mathrm{SWAP}} = \SI{400}{\nano\second}$ as
its QSL. The other two configurations are not capable of realizing SWAP gates.
However, since the SWAP gate, in contrast to the CNOT gate, can be realized with
global control fields, we believe it to be experimentally feasible and thus
include it as a viable gate in the QGS in Table~\ref{tab:gatesets}.

\begin{figure}[!tb]
  \centering
  \includegraphics{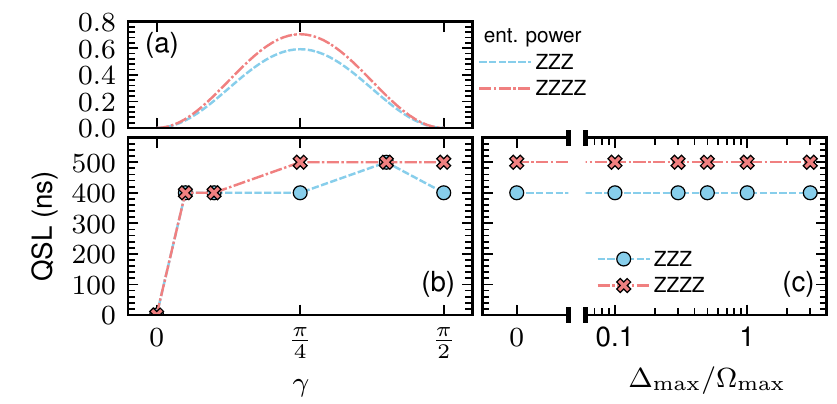}
  \caption{%
    Panel (a) shows the entanglement power~\cite{Linowski2020} of the three- and
    four-qubit constraint gates $\mathrm{ZZZ}(\gamma)$ and
    $\mathrm{ZZZZ}(\gamma)$, cf.\ Eq.~\eqref{eq:Zk}, as a function of $\gamma$.
    In contrast, panels (b) and (c) examine the QSLs for these gates under
    various conditions on neutral atoms. In panel (b), the dependence of the QSL
    on the parameters $\gamma$ is shown. In panel (c), the impact of
    $\Delta_{\rmmax}/\Omega_{\rmmax}$ is visualized. In the latter case, we have
    $\gamma = \pi/4$ and a fixed $\Omega_{\rmmax}$ while $\Delta_{\rmmax}$ is
    modified. The QSLs in panels (b) and (c) have been determined using the
    parameters and ``phase'' configuration described in Fig.~\ref{fig:qsl}.
  }
  \label{fig:ryd_duo}
\end{figure}

\subsubsection{Three- and four-qubit constraint gates}
So far, we have discussed the results for two-qubit gates in Fig.~\ref{fig:qsl}
(a)-(c). These gates and their respective QSLs have been used in determining the
quantum circuits and calculating the corresponding circuit run times in
Sec.~\ref{sec:main} --- especially for those circuits in the SGM discussed in
Sec.~\ref{subsec:standard}. In contrast, for QAOA circuits in the
PM~\cite{Lechner2020}, circuit representations without two-qubit gates exist,
e.g., when the required three- and four-qubit constraint gates
$\mathrm{ZZZ}(\gamma)$ and $\mathrm{ZZZZ}(\gamma)$, cf.\ Eq.~\eqref{eq:Zk}, are
available natively and thus must not be decomposed into single- and two-qubit
gates. In the following, we determine and discuss their QSLs.

It should first be noted that, from an algorithmic point of view, it is
irrelevant whether, e.g., $\mathrm{ZZZ}(\gamma)$ or $e^{\im \alpha}
\mathrm{ZZZ}(\gamma)$, with $\alpha$ some arbitrary phase, is realized in
experiments. The latter just changes the global phase of the quantum state
during circuit execution. The same holds for the $\mathrm{ZZZZ}(\gamma)$ gate.
In both cases, we may choose $\alpha$ arbitrarily but, for practical reasons,
choose it such that the states $\ket{\downarrow \downarrow \downarrow}$ and
$\ket{\downarrow \downarrow \downarrow \downarrow}$ do not acquire a phase from
the respective constraint gates. In the following, we therefore consider the
phase-shifted constraint gates
\begin{align} \label{eq:Zk_phase}
  e^{\im \gamma} \mathrm{ZZZ}(\gamma),
  \qquad
  e^{- \im \gamma} \mathrm{ZZZZ}(\gamma),
\end{align}
instead of the ones from Eq.~\eqref{eq:Zk}.

In the context of QAOA circuits in the PM, these constraint gates need to be
realized for various $\gamma$, cf.\ Eq.~\eqref{eq:qaoa_lhz}. However, since we
can not determine their QSL for each value of $\gamma$, we first analyze the
gate's entangling power~\cite{Linowski2020} as a function of $\gamma$ in
Fig~\ref{fig:ryd_duo} (a). We observe maximal entangling power for $\gamma
= \pi/4$ for both $\mathrm{ZZZ}(\gamma)$ and $\mathrm{ZZZZ}(\gamma)$ and thus
decide to first benchmark their QSLs for that particular value as we expect the
control problem in that case to be the hardest to solve and consequently the
QSLs to be the largest.

Figure~\ref{fig:qsl} (d) and (e) show the optimization results for the
phase-shifted constraint gates $\mathrm{ZZZ}(\gamma)$ and
$\mathrm{ZZZZ}(\gamma)$, cf.\ Eq.~\eqref{eq:Zk_phase}, for $\gamma = \pi/4$,
respectively. We use the same three configurations of control fields as for the
two-qubit gates in panels (a)-(c) and find all configurations to be capable of
implementing the constraint gates but observe the best performance for the
``parallel'' and ``phase'' configuration. While both configurations yield the
same QSLs of $T_{\rmqsl}^{\mathrm{ZZZ}} = \SI{400}{\nano\second}$ and
$T_{\rmqsl}^{\mathrm{ZZZZ}} = \SI{500}{\nano\second}$ for $\mathrm{ZZZ}(\gamma)$
and $\mathrm{ZZZZ}(\gamma)$, respectively, the ``phase'' configuration exhibits
the better convergence behavior. Like for the CZ gate in Fig.~\ref{fig:qsl} (a),
we observe every set of guess fields for this configuration to reliably converge
towards the same final error $\error_{T}$ for every $T$. Since the ``parallel''
configuration yields the same achievable errors as a function of gate time $T$
and uses by definition a different control strategy than the ``phase''
configuration, we believe to have reliably identified the QSLs for the
constraint gates. In terms of experimental feasibility, the ``phase''
configuration is advantageous as it requires fewer hardware and control
resources.

The fact that using $\phiu(t)$ as the only time-dependent control field suffices
for implementing $\mathrm{ZZZ}(\gamma)$ or $\mathrm{ZZZZ}(\gamma)$ originates
from considering the gates' phase-shifted versions of Eq.~\eqref{eq:Zk_phase}.
In detail, since the states $\ket{\downarrow \downarrow \downarrow}$ and
$\ket{\downarrow \downarrow \downarrow \downarrow}$ are the only states among
the $8$ or $16$ basis states of $\mathrm{ZZZ}(\gamma)$ or
$\mathrm{ZZZZ}(\gamma)$ that technically require non-zero $\Omd(t)$ and
$\phid(t)$ in order to be phase-configurable, we simply avoid this requirement
by considering the gates' phase-shifted version. For the remaining $7$ or $15$
basis states, which all have at least one atom initially in the $\ket{\uparrow}$
state, the phase $\phiu(t)$ together with a constant $\Omu(t) = \Omega_{\rmmax}$
is sufficient to correctly adjust all phases. The ``phase'' configuration thus
represents a hardware-efficient control scheme to realize fast
$\mathrm{ZZZ}(\gamma)$ and $\mathrm{ZZZZ}(\gamma)$ gates.

Our results furthermore reveal that the ``sequential'' configuration, which was
recently introduced in Ref.~\cite{PRL.128.120503} and designed to implement
high-fidelity $\mathrm{ZZZZ}(\gamma)$ gates, seems not to be ideal when it comes
to gate time as its configuration-specific QSL is roughly twice as long as the
QSLs for the other configurations. However, it should be noted that a comparison
of the protocol from Ref.~\cite{PRL.128.120503} with protocols at the QSL is not
a fair comparison. On the one hand, the control scheme of
Ref.~\cite{PRL.128.120503} is based on adiabaticity --- a regime that we are far
away from in our numerical calculations. On the other hand, while the parameter
$\gamma$ is a tunable variable in the adiabatic control scheme, the results in
Fig.~\ref{fig:qsl} (d) and (e) are only valid for $\gamma = \pi/4$. If gates
with a different $\gamma$ are required, one explicitly needs to optimize control
fields for that purpose. While it is beyond the scope of this work to examine
whether there exists an analytical control scheme with configurable $\gamma$ at
the QSL, in the following, we nevertheless provide an analysis of the QSLs for
$\mathrm{ZZZ}(\gamma)$ and $\mathrm{ZZZZ}(\gamma)$ beyond $\gamma = \pi/4$. In
detail, after having identified the ``phase'' configuration as the most reliable
configuration to determine a gate's QSL, we provide the QSLs for other $\gamma$
values in Fig.~\ref{fig:ryd_duo} (b). We find the QSLs to be almost constant for
$\gamma>0$ and only zero for $\gamma = 0$, in which case $\mathrm{ZZZ}(\gamma)$
and $\mathrm{ZZZZ}(\gamma)$ coincide with the identity operation.
Interestingly, we do not observe a decrease in the QSLs for $\gamma = \pi/2$ in
which case the constraint gates are no longer entangling, cf.\
Fig.~\ref{fig:ryd_duo} (a), and should therefore theoretically be implementable
with local operations only. We suspect that we do not see a decrease of the QSLs
since we do not consider any control fields for local operations in
Hamiltonian~\eqref{eq:ham_ryd} and thus need to implement the local gates by
means of the Rydberg levels.

We moreover analyze the dependence of the QSLs on the ratio $\Delta_{\rmmax}
/ \Omega_{\rmmax}$ in Fig.~\ref{fig:ryd_duo} (c). Surprisingly, we observe the
QSLs for both $\mathrm{ZZZ}(\gamma)$ and $\mathrm{ZZZZ}(\gamma)$ to be
independent on this ratio and find $\Delta_{\rmmax} = 0$ to be a viable option.

In general, we observe that the QSLs for $\mathrm{ZZZ}(\gamma)$ and
$\mathrm{ZZZZ}(\gamma)$ are only slightly larger than those of the two-qubit
gates. In view of quantum circuits for PM-QAOA, where such constraint gates are
required, it is thus advantageous to have these gates natively available, since
their representation via single- and two-qubit gates~\cite{Lechner2020} consumes
significantly more time. This effect can be seen in Fig.~\ref{fig:comp} (c),
where the plus signs illustrate the data for constraint gates expanded in
single- and two-qubit gates and the crosses for the usage of native constraint
gates. One possible explanation for the short QSLs for $\mathrm{ZZZ}(\gamma)$
and $\mathrm{ZZZZ}(\gamma)$ compared to those of the two-qubit gates might be
that the permutation symmetry of atoms within the pseudo-2D
architecture~\cite{PRL.128.120503}, i.e., $V_{nm} = V$, matches the permutation
symmetry of the gate operation itself.

\begin{figure*}[!tb]
  \centering
  \includegraphics{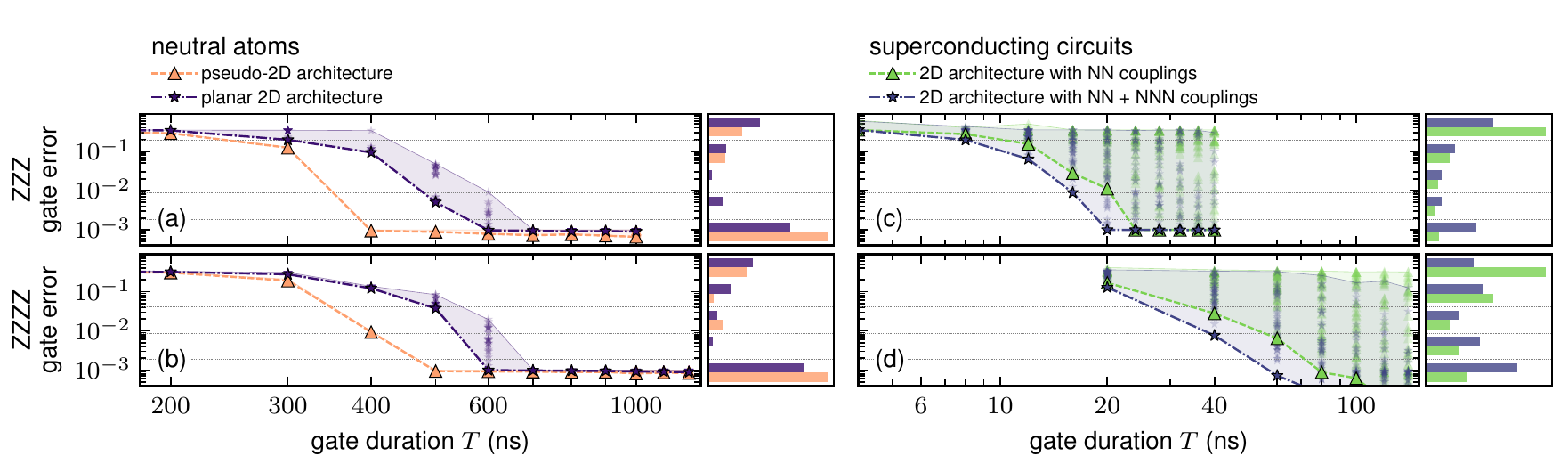}
  \caption{%
    Overview of different QSLs similar to Fig.~\ref{fig:qsl} but exclusively for
    the three- and four-qubit constraint gates $\mathrm{ZZZ}(\gamma)$ and
    $\mathrm{ZZZZ}(\gamma)$, cf.\ Eq.~\eqref{eq:Zk}. The left column shows
    results for neutral atoms, obtained using the ``phase'' configuration of
    Fig.~\ref{fig:qsl}, for the pseudo-2D architecture (triangles) and an
    actual, planar 2D architecture (stars). The right column compares the
    results for superconducting circuits using the ``no-X'' configuration from
    Fig.~\ref{fig:qsl}. The data correspond to the physical architecture of
    Ref.~\cite{Arute2019} where only nearest neighbor (NN) couplings between
    qubits are present (triangles) and where diagonal couplings, i.e.,
    next-to-nearest neighbor (NNN) couplings, are added (stars).
  }
  \label{fig:qsl2D}
\end{figure*}

At last, we therefore examine the impact of the pseudo-2D architecture onto the
constraint gates $\mathrm{ZZZ}(\gamma)$ and $\mathrm{ZZZZ}(\gamma)$. For the
three- and four-qubit constraint gates, the change to an actual, planar 2D
architecture implies that while the couplings between nearest neighbors remain
$V$, diagonal couplings, i.e., couplings between next-to-nearest neighbors or,
in other words, qubits on opposite edges of a $2 \times 2$ square plaquette, are
replaced by $V/8$. The stars in Fig.~\ref{fig:qsl2D} (a) and (b) show the
corresponding results for $\mathrm{ZZZ}(\gamma)$ and $\mathrm{ZZZZ}(\gamma)$
gates and $\gamma = \pi/4$, respectively. While their QSLs within the pseudo-2D
architecture are $T_{\rmqsl}^{\mathrm{ZZZ}} = \SI{400}{\nano\second}$ and
$T_{\rmqsl}^{\mathrm{ZZZZ}} = \SI{500}{\nano\second}$, they become
$T_{\rmqsl,\mathrm{2D}}^{\mathrm{ZZZ}} = T_{\rmqsl,\mathrm{2D}}^{\mathrm{ZZZZ}}
= \SI{600}{\nano\second}$ in the actual, planar 2D architecture. Interestingly,
this corresponds only to a relatively small increase in the QSLs for both gates.
A possible explanation might be that the gate speed for neutral atoms is
primarily determined by the maximal Rabi frequency $\Omega_{\rmmax}$, which is
identical for both examples, and not so much by the interatomic interaction
strength, which is different for both architectures.

Although we believe the pseudo-2D architecture to be viable in experiments due
to the great flexibility to arrange neutral atoms~\cite{Barredo2018}, we
nevertheless take the QSLs for the actual, planar 2D architecture to be the
reference gate times within the QGS in Table~\ref{tab:gatesets} and
Sec.~\ref{sec:main}. However, recall that the constraint gates are --- within
our study --- only relevant for the QAOA circuits in the PM, cf.\
Fig.~\ref{fig:comp} (c). In order to nevertheless allow for a comparison of the
run times in the actual, planar 2D architecture (orange crosses) with those
using the pseudo-2D architecture, we add the latter as orange pentagons to
Fig.~\ref{fig:comp} (c).

\subsection{Superconducting circuits}
\label{subsec:sc}
Similar to neutral atoms in Sec.~\ref{subsec:ryd}, we now determine and analyze
the QSLs for the same quantum gates but for superconducting circuits. The
available control knobs to implement these gates are the tunable qubit
frequencies $\omega_{n}(t)$, the tunable coupling strength $g_{nm}(t)$ between
qubits and the (auxiliary) X-type local control fields $\bar{\Omega}_{n, \rmre}
(t)$ and $\bar{\Omega}_{n, \rmim} (t)$ in Hamiltonian~\eqref{eq:ham_sc'}. In
order to remain experimentally realistic, we take parameters from
Ref.~\cite{Arute2019}. To this end, we single out a $2 \times 2$ plaquette
consisting of four qubits from the generally larger 2D architecture. We take the
qubit frequencies and their tunable range to be given by $\SI{6700}{\mega\hertz}
\lesssim \omega_{n}(t) \lesssim \SI{7100}{\mega\hertz}$ and their
anharmonicities given by $\alpha_{n} \approx \SI{200}{\mega\hertz}$, with exact
values depending on $n$. The tunable coupling strength is given by
$-\SI{40}{\mega\hertz} \leq g_{nm}(t) \leq \SI{5}{\mega\hertz}$, as reported in
Ref.~\cite{Arute2019}. Moreover, we assume the (auxiliary) X-type control fields
$\bar{\Omega}_{n, \rmre} (t)$ and $\bar{\Omega}_{n, \rmim} (t)$, which encode
the physical X-type control fields $\Omega_{n}(t)$ and their tunable driving
frequencies $\bar{\omega}_{n}(t)$, to satisfy $-\SI{50}{\mega\hertz} \leq
\bar{\Omega}_{n, \rmre} (t), \bar{\Omega}_{n, \rmim} (t) \leq
\SI{50}{\mega\hertz}$.

\subsubsection{Two-qubit gates}
Figure~\ref{fig:qsl} (f) shows results for a CZ gate on superconducting circuits
using three different configurations of control fields --- different, of course,
from those used for neutral atoms. In the ``full'' configuration (circles), all
available control fields, $\omega_{n}(t), g_{nm}(t), \bar{\Omega}_{n, \rmre}
(t)$ and $\bar{\Omega}_{n, \rmim} (t)$, are time-dependent and being optimized.
In the ``no-X'' configuration (crosses), only $\omega_{n}(t)$ and $g_{nm}(t)$
are optimized while the X-type control fields are set to zero, $\bar{\Omega}_{n,
\rmre} (t) = \bar{\Omega}_{n, \rmim} (t) = 0$. At last, in the ``interaction''
configuration only $\omega_{n}(t)$ is time-dependent and optimized while
$g_{nm}(t) = -\SI{40}{\mega\hertz} = g_{\rmmax}$ is set to its maximum magnitude
and $\bar{\Omega}_{n, \rmre} (t) = \bar{\Omega}_{n, \rmim} (t) = 0$. For the CZ
gate in Fig.~\ref{fig:qsl} (f), we observe all three configurations to indicate
the same QSL of $T_{\rmqsl}^{\mathrm{CZ}} = \SI{10}{\nano\second}$. In terms of
convergence behavior, the ``interaction'' configuration shows the best
performance as indicated by the probability density on the right side of
Fig.~\ref{fig:qsl} (f). In general, the three configurations show slightly worse
convergence behavior than the three configurations for the neutral atoms, cf.\
Fig.~\ref{fig:qsl} (a). Nevertheless, since all three configurations indicate
the same QSL and, in general, yield the same achievable error $\error_{T}$
depending on $T$, we believe our method to determine the QSL to yield reliable
results also for superconducting circuits and conjecture
$T_{\rmqsl}^{\mathrm{CZ}} = \SI{10}{\nano\second}$ to be the fundamental QSL for
CZ gates. While there are reference implementations for CZ gates on similar
architectures with tunable couplers~\cite{PRL.113.220502, PRX.11.021058,
PRL.125.240503, PRApplied.14.024070}, none of these architectures matches our
architecture and parameter regime. Hence, we compare our QSL to the gate time of
the fastest two-qubit gate, the Sycamore gate, reported in
Ref.~\cite{Arute2019}. We find $T_{\rmqsl}^{\mathrm{CZ}}
= \SI{10}{\nano\second}$ to be slightly faster than $T_{\rmlit}^{\mathrm{Syc.}}
= \SI{12}{\nano\second}$~\cite{datasheet}.

In Fig.~\ref{fig:qsl} (g), the results for a CNOT gate are shown using the same
three configuration as for the CZ gate. We find only the ``full'' configuration
to be capable of realizing a CNOT gate, yielding the QSL
$T_{\rmqsl}^{\mathrm{CNOT}} = \SI{14}{\nano\second}$, while the other two
configuration are not capable of it. Among the available control knobs, the
X-type control fields $\bar{\Omega}_{n, \rmre} (t)$ and $\bar{\Omega}_{n, \rmim}
(t)$ are crucial for a CNOT gate to be feasible. For the SWAP gate, we again
find all three configuration to converge, cf.\ Fig.~\ref{fig:qsl} (h), with the
``no-X'' and ``interaction'' configuration showing the best convergence
behavior. We find a QSL of $T_{\rmqsl}^{\mathrm{SWAP}} = \SI{12}{\nano\second}$.

\subsubsection{Three- and four-qubit constraint gates}
We now turn towards the three- and four-qubit constraint gates
$\mathrm{ZZZ}(\gamma)$ and $\mathrm{ZZZZ}(\gamma)$. However, note that in the
following and in contrast to neutral atoms, we do not consider their
phase-shifted versions, cf.\ Eq.~\eqref{eq:Zk_phase}, but their original
versions, cf.\ Eq.~\eqref{eq:Zk}.

Figure~\ref{fig:qsl} (i) and (j) show the results for the constraint gates
$\mathrm{ZZZ}(\gamma)$ and $\mathrm{ZZZZ}(\gamma)$, respectively. We use the
same three configurations of control fields as for the two-qubit gates of panels
(f)-(h). We observe the ``interaction'' configuration to have the best
convergence behavior while the ``no-X'' configuration gives in both cases rise
to the shortest QSLs of $T_{\rmqsl}^{\mathrm{ZZZ}} = \SI{24}{\nano\second}$ and
$T_{\rmqsl}^{\mathrm{ZZZZ}} = \SI{80}{\nano\second}$ for $\mathrm{ZZZ}(\gamma)$
and $\mathrm{ZZZZ}(\gamma)$, respectively. Both QSLs have been determined for
$\gamma = \pi/4$. The QSL for $\mathrm{ZZZ}(\gamma)$ has thereby been confirmed
independently by both the ``full'' and ``no-X'' configurations as both yield
almost identical achievable errors $\error_{T}$ as a function of gate time $T$.
We thus assume the QSL $T_{\rmqsl}^{\mathrm{ZZZ}} = \SI{24}{\nano\second}$ to be
well backed up. The situation is different for $\mathrm{ZZZZ}(\gamma)$, for
which we observe very different convergence behaviors for the three
configurations, cf.\ Fig.~\ref{fig:qsl} (j). While the ``no-X'' configuration
yields the shortest QSL of $T_{\rmqsl}^{\mathrm{ZZZZ}} = \SI{80}{\nano\second}$,
the ``full'' configuration shows slightly better performance for $T
< T_{\rmqsl}^{\mathrm{ZZZZ}}$, which might suggest that even shorter gate
protocols for $\mathrm{ZZZZ}(\gamma)$ may exist but our method did not find them
due, e.g., limited numbers of guess fields for exploring the control landscape.
In the following, as well as for the calculation of circuit run times in
Sec.~\ref{sec:main}, we nevertheless assume $T_{\rmqsl}^{\mathrm{ZZZZ}}
= \SI{80}{\nano\second}$ to be the QSL for the $\mathrm{ZZZZ}(\gamma)$ gate as
it is the fastest gate time $T$ among the three configurations of control fields
for which Krotov's method was able to find a solution with $\error_{T} \leq
\error_{\rmmax}$.

Interestingly, while we observe the QSLs for $\mathrm{ZZZ}(\gamma)$ and
$\mathrm{ZZZZ}(\gamma)$ to be almost identical for neutral atoms and only
slightly longer than the QSLs for the two-qubit gates, we observe the same only
for the $\mathrm{ZZZ}(\gamma)$ gate for superconducting circuits. The
$\mathrm{ZZZZ}(\gamma)$ gate has a much longer QSL compared to the other QSLs on
that platform. To rigorously decide whether this is due to the non-ideal
convergence behavior observed in Fig.~\ref{fig:qsl} (j) or has some deeper
physical origin is beyond the scope of this study. In an attempt to tackle this
question nevertheless, we consider the scenario of having additional diagonal
couplings, i.e., next-to-nearest neighbor couplings, among the transmons in the
superconducting circuit architecture. For Hamiltonian~\eqref{eq:ham_sc_int},
this implies adding two additional rows with couplings $g_{13}(t)$ and
$g_{24}(t)$ in the same form as the already present couplings $g_{12}(t),
g_{23}(t), g_{34}(t)$ and $g_{41}(t)$. This scenario is inspired by the
pseudo-2D architecture for neutral atoms, which also exhibits identical nearest
neighbor (NN) and next-to-nearest neighbor (NNN) couplings and where having
these couplings is advantageous. Figure~\ref{fig:qsl2D} (c) and (d) show the
results for the two cases with only NN couplings (triangles) and with NN plus
NNN couplings (stars). Besides observing much better convergence properties for
the latter case, we also obtain improved QSLs of
$T_{\rmqsl,\mathrm{NNN}}^{\mathrm{ZZZ}} = \SI{20}{\nano\second}$ and
$T_{\rmqsl,\mathrm{NNN}}^{\mathrm{ZZZZ}} = \SI{60}{\nano\second}$ for the
$\mathrm{ZZZ}(\gamma)$ and $\mathrm{ZZZZ}(\gamma)$ gate, respectively. However,
despite improvements in this scenario, the QSL of the $\mathrm{ZZZZ}(\gamma)$
gate does not get close to the QSL of the $\mathrm{ZZZ}(\gamma)$ gate as it does
for neutral atoms. The presence of NNN couplings might therefore be just
a partial explanation of the QSL differences for the constraint gates between
neutral atoms and superconducting circuits. Despite the scenario with NNN
couplings not reflecting the actual architecture for superconducting circuits,
we nevertheless add the circuit run times for the QAOA circuits in the PM using
these faster constraint gates to Fig.~\ref{fig:comp} (c) for reference purposes
as purple pentagons.

\section{Conclusions}
\label{sec:conclusions}
In this study, we have determined the QSLs for several common two-qubit and two
specific multi-qubit quantum gates for two promising quantum computing platforms
that allow for a 2D arrangement of qubits --- neutral atoms and superconducting
circuits. We have used OCT to determine the QSLs, as it provides a generally
applicable tool that warrants a fair comparison of both platforms.

On the level of individual quantum gates, our study allows assessing how close
gate protocols from the literature are compared to their fundamental QSLs or, in
other words, how much can (theoretically) still be gained in time if known gate
protocols are replaced by numerically optimized ones. We find the QSLs for all
investigated two-qubit gates, encompassing CNOT, CZ and SWAP, to be very similar
within each platform and close to the reference gate times for a CZ gate in case
of neutral atoms~\cite{PRL.123.170503} and a Sycamore gate in case of
superconducting circuits~\cite{Arute2019}.

On the level of quantum algorithms, our study has moreover allowed us to
determine the ``QSLs'' for entire quantum circuits. To this end, we have assumed
a 2D grid architecture for both platforms and qubit connectivity that allows
physical quantum gates only between neighboring qubits. However, we have assumed
these gates to be executable at the QSL. This has allowed us to calculate the
circuit run times at the ``QSL'' for two paradigmatic quantum algorithms --- the
QFT and a single step of the QAOA. We find that the corresponding weighted
circuit run times scale comparably with respect to the system size. Furthermore,
we observe this to be independent of the chosen gate set used to translate the
quantum algorithms into executable quantum circuits, i.e., independent of
whether the SGS or QGS is used. We observe platform-independently that the QGS
yields circuit run times and gate counts that are roughly half compared to those
in the SGS. On the one hand, this demonstrates that further speedup of circuit
run times is theoretically possible on both platforms. On the other hand, it
also shows that both platforms perform equally well when running prototypical
quantum circuits using typical, present-day NISQ hardware.

Besides a representation of the quantum circuits in the SGM, we have also
explored the representation of the same quantum algorithms in the
PM~\cite{Lechner2015, Lechner2020, PRL.129.180503}. We observe a reduction in
circuit run times as well as in gate counts in most cases. This reduction comes
at the expense of requiring more physical qubits but without the need to change
the geometrical layout or the control hardware. In this context, we want to
specifically emphasize the circuit run times of a single QAOA step in the PM.
Compared to its representation in the SGM, it offers a constant circuit depth
independent on the problem complexity but requires the implementation of local
three- and four-qubit constraint gates~\cite{Lechner2020}. For superconducting
circuits we find their gate times at the QSL to be roughly similar to the run
times of their decompositions into single- and two-qubit gates. In contrast, for
neutral atoms we find the direct implementation of the constraint gates to be
only slightly slower than any single two-qubit gate and especially much faster
than their decompositions into single- and two-qubit gates. We nevertheless
observe for both platforms run times for a single QAOA step on the order of
$2-4\%$ of the platform's intrinsic coherence time. While this corresponds to an
improvement of one order of magnitude for a problem size with $N=9$ logical
qubits, this grows to an improvement of three orders of magnitude for $N=121$.
Since the gate counts, when the native implementations of the constraint gates
are used, are also smaller in the PM compared to the SGM, we believe the PM to
be advantageous in terms of the described resources for a single QAOA step. It
minimizes errors due to finite coherence time and allows for large-depth QAOA.
However, we want to emphasize that deeper PM-QAOA circuits are in general
necessary to reach similar success probabilities compared to lower-depth
SGM-QAOA implementations.

While we want to emphasize that the feasibility of a quantum circuit depends on
more than just its run time and gate count, our benchmark study demonstrates
that at least these two factors can be theoretically further improved using
optimized gate protocols. This holds for both neutral atoms and superconducting
circuits. 

\begin{acknowledgments}
  We would like to thank Hannes Pichler and Kilian Ender for helpful discussions
  and Michael Fellner for help with the parity QFT circuits. This work was
  supported by the Austrian Science Fund (FWF) through a START grant under
  Project No. Y1067-N27 and I 6011. This research was funded in whole, or in
  part, by the Austrian Science Fund (FWF) SFB BeyondC Project No. F7108-N38.
  For the purpose of open access, the author has applied a CC BY public
  copyright licence to any Author Accepted Manuscript version arising from this
  submission. This project was funded within the QuantERA II Programme that has
  received funding from the European Union's Horizon 2020 research and
  innovation programme under Grant Agreement No. 101017733. 
\end{acknowledgments}

\appendix


\section{Quantum circuits for QFT and QAOA}
\label{app:circs}

\subsection{Standard gate model}
\label{app:circs:standard}
The quantum Fourier transform (QFT) is a key ingredient in Shor's algorithm for
integer factorization~\cite{SIAM.26.1484} and thus a prototypical application
for quantum computers. While a typical circuit representation of a QFT contains
exclusively Hadamard gates, $H$, and controlled-phase gates, $R_{n}
= \rmdiag\{1,1,1,\exp\{2 \pi \im / 2^{n}\}\}$, a more efficient representation
with fewer gates and lower circuit depths can be constructed using SWAP
gates~\cite{Holmes2020}.

The quantum approximate optimization algorithm (QAOA) aims at finding
approximate solution to combinatorial optimization problems~\cite{Farhi2014}.
For instance, let us consider the task to find the ground state of the $N$ qubit
spin glass Hamiltonian
\begin{align} \label{eq:Hz}
  \op{H}_{\rmz}
  =
  \sum_{\substack{n,m=1 \\ n < m}}^{N}
  J_{nm} \op{\sigma}_{\rmz}^{(n)} \op{\sigma}_{\rmz}^{(m)},
\end{align}
where $J_{nm}$ denotes the interaction strength between qubits $n$ and $m$ and
$\op{\sigma}_{\rmz}^{(i)}$ the Pauli-z operator on qubit $i$. The QAOA allows to
find an approximate solution, i.e., an approximate ground state for
Hamiltonian~\eqref{eq:Hz}, by applying the procedure
\begin{align} \label{eq:qaoa}
  \ket{\psi_{\rmout}}
  =
  \prod_{k=1}^{p}
  e^{- \im \alpha_{k} \op{H}_{\rmx}}
  e^{- \im \beta_{k}  \op{H}_{\rmz}}
  \ket{\psi_{\rmin}},
  \qquad
  \op{H}_{\rmx}
  =
  \sum_{n=1}^{N} \op{\sigma}_{\rmx}^{(n)},
\end{align}
where $\ket{\psi_{\rmin}}$ is the ground state of the so-called mixing
Hamiltonian $\op{H}_{\rmx}$ and $\alpha_{k}, \beta_{k} \in [0, 2\pi)$ are angles
--- physically corresponding to evolution times --- that are iteratively
optimized via a classical, closed-loop feedback optimization and the energy
expectation value of $\ket{\psi_{\rmout}}$ being the objective to minimize. The
number of steps is denoted by $p$. A single step in the QAOA is thus given by
the application of the spin glass or problem Hamiltonian $\op{H}_{\rmz}$
followed by the mixing Hamiltonian $\op{H}_{\rmx}$. While the latter corresponds
to a parallel application of $\alpha_{k}$-dependent single-qubit $X$ rotations,
$R_{\rmx}^{k}$, in the associated quantum circuit --- and is thus negligible
time-wise --- the circuit implementation of the former requires multiple CNOT
gates as well as single-qubit phase gates, $R_{\rmz}^{nm} = \exp\{-\im \beta_{k}
J_{nm} \op{\sigma}_{\rmz}\}$, containing information about $J_{nm}$ and
$\beta_{k}$.

\subsection{Parity mapping}
\label{app:circs:parity}
In the SGM, as described in Appendix~\ref{app:circs:standard}, every logical
qubit is given by exactly one physical qubit and gates on logical qubits are
equivalent to gates on physical qubits. This is different for the PM, where $K
> N$ physical qubits are required for a problem of $N$ logical qubits and gates
on logical qubits become different gates or even gate sequences on the physical
qubits. However, due to the arrangement of physical qubits according to the PM,
all gates between physical qubits are strictly local and thus require only
nearest-neighbor connectivity.

For the QFT, we need $K = N(N+1)/2$ physical qubits in the PM and find that
Hadamard gates, $H$, on logical qubits become equivalent to several single- and
two-qubit gates on neighboring physical qubits. In contrast, the logical
two-qubit controlled-phase gates, $R_{n}$, between any pair for logical qubits
is given by exactly three parallel single-qubit gates on physical
qubits~\cite{PRL.129.180503, PRA.106.042442}. Hence, while the logical Hadamard
gates require more resources in the PM, the logical controlled-phase gates
require significantly less resources --- especially for those gates where the
logical qubits are far away from each other.

For the QAOA, we need $K = N(N-1)/2$ physical qubits in the PM and
Eq.~\eqref{eq:qaoa} becomes~\cite{Lechner2020}
\begin{align} \label{eq:qaoa_lhz}
  \ket{\psi_{\rmout}}
  =
  \prod_{k=1}^{p}
  e^{- \im \alpha_{k} \op{H}_{\rmx}^{\rmphys}}
  e^{- \im \beta_{k}  \op{H}_{\rmz}^{\rmphys}}
  e^{- \im \gamma_{k} \op{H}_{\rmc}}
  \ket{\psi_{\rmin}},
\end{align}
where
\begin{align} \label{eq:Hx_Hz_lhz}
  \op{H}_{\rmx}^{\rmphys}
  =
  \sum_{k=1}^{K} \tilde{\op{\sigma}}_{\rmx}^{(k)},
  \qquad
  \op{H}_{\rmz}^{\rmphys}
  =
  \sum_{k=1}^{K} \tilde{J}_{k} \tilde{\op{\sigma}}_{\rmz}^{(k)}
\end{align}
are the modified mixing and spin glass Hamiltonians in the PM, respectively.
The local field strengths $\tilde{J}_{k}$ run over the $N(N-1)/2$ interactions
$J_{nm}$, cf.\ Eq.~\eqref{eq:Hz}, and $\tilde{\op{\sigma}}_{\rmz}^{(k)}$ encodes
the parity of the corresponding two-qubit interaction $\op{\sigma}_{\rmz}^{(n)}
\op{\sigma}_{\rmz}^{(m)}$~\cite{Lechner2015}. In order to solve optimization
problems using the PM, we additionally need to constraint the dynamics to the
$2^{N-1}$ dimensional subspace within the $2^{K}$ dimensional physical Hilbert
space that corresponds to the $2^{N-1}$ eigenstates of Eq.~\eqref{eq:Hz} that
have unique eigenvalues~\footnote{Note that every eigenstate of
Eq.~\eqref{eq:Hz} appears at least twice due to a global spin flip symmetry}.
Hence, since not every eigenstates of $\op{H}_{\rmz}^{\rmphys}$ has a logical
counterpart in $\op{H}_{\rmz}$, it needs to be energetically penalized as it
would not be a valid solution to the optimization problem. This is achieved by
realizing $C=K-N+1$ local three- and four-qubit constraints
via~\cite{Lechner2015}
\begin{align} \label{eq:Hc_lhz}
  \op{H}_{\rmc}
  =
  \sum_{c=1}^{C}
  \tilde{\op{\sigma}}_{\rmz}^{(k_{1})}
  \tilde{\op{\sigma}}_{\rmz}^{(k_{2})}
  \tilde{\op{\sigma}}_{\rmz}^{(k_{3})}
  \left(\tilde{\op{\sigma}}_{\rmz}^{(k_{4})}\right),
\end{align}
where ``local'' refers to $k_{1}, \dots, k_{4}$ being nearest-neighbor physical
qubits. For every QAOA step in Eq.~\eqref{eq:qaoa_lhz}, this requires $C$
constraint gates of the form (neglecting tildes and indices)
\begin{subequations} \label{eq:Zk}
\begin{align}
  \mathrm{ZZZ}(\gamma)
  &=
  \exp\left\{%
    - \im \gamma
    \op{\sigma}_{\rmz} \op{\sigma}_{\rmz} \op{\sigma}_{\rmz}
  \right\},
  \\
  \mathrm{ZZZZ}(\gamma)
  &=
  \exp\left\{%
    - \im \gamma
    \op{\sigma}_{\rmz} \op{\sigma}_{\rmz} \op{\sigma}_{\rmz} \op{\sigma}_{\rmz}
  \right\}
\end{align}
\end{subequations}
with $\gamma$ the effective constraint strengths, which --- like in the original
QAOA scheme of Eq.~\eqref{eq:qaoa} --- are optimized in a classical, closed-loop
feedback optimization. The gates in Eq.~\eqref{eq:Zk} can be either realized
directly~\cite{PRL.128.120503} or by decomposing them into single- and two-qubit
gates, e.g., by using four or six CNOT gates plus one $\gamma$-dependent
single-qubit phase-gate for $\mathrm{ZZZ}(\gamma)$ or $\mathrm{ZZZZ}(\gamma)$,
respectively~\cite{Lechner2020}.

It is important to note that the constraint gates are the only multi-qubit gates
in Eq.~\eqref{eq:qaoa_lhz}. Independent of $N$, their implementation can be
parallelized with at most nine~\cite{PRL.128.120503} or four~\cite{Lechner2020}
consecutive layers of constraint gates for neutral atoms or superconducting
circuits, respectively. However, note that shallower circuits may be feasible
by now~\cite{Unger2022}. The difference between the two platforms originates
from their different qubit-qubit coupling mechanism. The tunable coupler
architecture of superconducting circuits~\cite{Arute2019} allows to switch off
the coupling between any pair of qubits. As a consequence, all constraint
plaquettes that do not share a common qubit, i.e., next-to-nearest neighbor
plaquettes, can be implemented in parallel. A single QAOA step thus requires at
most four layers of constraint gates to realize all of them. Neutral atoms, in
contrast, interact via their Rydberg levels and therefore require additional
spatial separation between atoms that are in their Rydberg levels but are not
supposed to interact, i.e., to suppress unwanted interactions. Assuming that
a single line of atoms in non-Rydberg levels suffices as a buffer between
plaquettes that are to be implemented in parallel, this corresponds to two lines
of plaquettes as a buffer in each spatial direction. This yields a maximal
number of nine layers of constraint gates. Despite the platform-dependent
differences, all quantum circuits corresponding to Eq.~\eqref{eq:qaoa_lhz} have
constant circuit depth.

\section{Krotov's method for quantum optimal control}
\label{app:krotov}
Krotov's method~\cite{AutomRemContr.60.1427} is an iterative, gradient-based
optimization algorithm for time-continuous control fields featuring a build-in
monotonic convergence~\cite{JCP.136.104103}. To achieve the latter, Krotov's
method requires a specific choice of the total optimization functional $J$, cf.\
Eq.~\eqref{eq:J}. In detail, while the error measure $\error_{T}$ at final time
$T$ remains the relevant figure of merit that we want to minimized, Krotov's
method achieves its minimization only indirectly by minimizing the total
functional $J$ where the time-dependent running costs $J_{t}$ are give
by~\cite{PRA.68.062308}
\begin{align} \label{eq:Jt}
  J_{t}\left[\{\psi_{l}(t)\}, \{\pulse_{k}(t)\}, t\right]
  =
  \sum_{k} \frac{\lambda_{k}}{S_{k}(t)}
  \left(\pulse_{k}(t) - \pulse_{k}^{\rmref}(t)\right)^{2},
\end{align}
where $\pulse_{k}^{\rmref}(t)$ is a reference field for the control field
$\pulse_{k}(t)$ that is to be optimized, $S_{k}(t) \in (0,1]$ is a shape
function and $\lambda_{k} > 0$ a numerical parameter. With the choice of
Eq.~\eqref{eq:Jt}, the update equation for field $\pulse_{k}(t)$
becomes~\cite{JCP.136.104103}
\begin{widetext}
  \begin{align} \label{eq:update}
    \pulse_{k}^{(i+1)}(t)
    =
    \pulse_{k}^{\rmref}(t)
    +
    \frac{S_{k}(t)}{\lambda_{k}} \ip\left\{%
      \sum_{l} \Braket{%
        \chi^{(i)}_{l}(t) |
        \frac{%
          \partial \op{H}\left[\left\{\pulse_{k'}\right\}\right]
        }{%
          \partial \pulse_{k}
        } \Big|_{\{\pulse^{(i+1)}_{k'}(t)\}}
        | \psi^{(i+1)}_{l}(t)
      }
    \right\},
  \end{align}
\end{widetext}
where $\ket{\psi_{l}^{(i+1)}(t)}$ are forward-propagated states and solutions to
the Schr\"{o}dinger equations
\begin{subequations} \label{eq:fw_states}
  \begin{equation}
    \frac{\dd}{\dd t} \Ket{\psi^{(i+1)}_{l}}(t)
    =
    - \im \op{H}^{(i+1)}(t)
    \Ket{\psi^{(i+1)}_{l}(t)}
  \end{equation}
with boundary conditions given by the initial states
  \begin{equation} \label{eq:fw_states_bound}
    \Ket{\psi^{(i+1)}_{l}(0)}
    =
    \Ket{\psi_{l}(0)}.
  \end{equation}
\end{subequations}
In contrast, $\ket{\chi^{(i)}_{l}(t)}$ are backward-propagated co-states and
solutions to the equations
\begin{subequations} \label{eq:bw_states}
  \begin{equation}
    \frac{\dd}{\dd t} \Ket{\chi^{(i)}_{l}(t)}
    =
    \im \op{H}^{(i)}(t)
    \Ket{\chi^{(i)}_{l}(t)}
  \end{equation}
with boundary conditions
  \begin{equation} \label{eq:bw_states_bound}
    \Ket{\chi^{(i)}_{l}(T)}
    =
    - \frac{%
      \partial \error_{T}
    }{%
      \partial \Bra{\psi_{l}}
    }
    \Bigg|_{\left\{%
        \Psi_{l}^{(i)}(T)
      \right\}}.
  \end{equation}
\end{subequations}
The superscripts $i$ and $i+1$ in Eqs.~\eqref{eq:update}-\eqref{eq:bw_states}
indicate whether the corresponding quantity is calculated using the ``old''
fields from iteration $i$ or the updated fields from iteration $i+1$,
respectively.

In order to turn Eq.~\eqref{eq:update} into a proper update equation, the
reference field $\pulse_{k}^{\rmref}(t)$ is taken to be the field
$\pulse_{k}^{(i)}(t)$ from the previous iteration in which case the second term
of the left-hand side of Eq.~\eqref{eq:update} becomes its update. This choice
causes the time-dependent costs $J_{t}$, cf.\ Eq.~\eqref{eq:Jt}, to gradually
vanish as the iterative procedure converges. Hence, the error-measure
$\error_{T}$ at final time $T$ becomes the dominant term within the total
optimization functional $J$, cf.\ Eq.~\eqref{eq:J}, and is thus predominantly
minimized. Equation~\eqref{eq:update} also reveals that while $\lambda_{k}$ can
be used to control the general size of the update, $S_{k}(t)$ can be used to
suppress updates at certain times.

We refer to Ref.~\cite{JCP.136.104103} for a more detailed introduction of
Krotov's method and to Ref.~\cite{SciPostPhys.7.6.080} for a detailed discussion
about its numerical implementation.

\section{Field generation and parametrization}
\label{app:field}
In this Appendix, we specify some of the technicalities for the method described
in Sec.~\ref{subsec:method}.

We generate randomized, albeit smooth, guess fields via the
formula~\cite{Filip2019}
\begin{align} \label{eq:f}
  f(t)
  =
  a_{0} + \sqrt{2} \sum_{j=1}^{m} \left[
    a_{j} \cos\left(\frac{2 \pi j t}{t_{1} - t_{0}}\right)
    +
    b_{j} \sin\left(\frac{2 \pi j t}{t_{1} - t_{0}}\right)
  \right]
\end{align}
where $a_{0}, a_{j}, b_{j}$ are chosen randomly from a normal distribution
$N(\mu, \sigma)$ with center $\mu=0$ and variance $\sigma = 1/(2 m + 1)$. The
integer $m$ thereby not only determines the number of frequency components in
$f(t)$ but also the frequencies of each component. Since we expect frequencies
on the same time scale than those contained in the Hamiltonian to be of greater
relevance for finding solutions, we chose $m$ randomly from an interval matching
each Hamiltonian's frequencies.

Besides the generation of randomized guess fields, we also specify the internal
parametrization of the fields. To this end, note that some of the physical or
auxiliary control fields in Eqs.~\eqref{eq:ham_ryd} or \eqref{eq:ham_sc'} are
experimentally limited in range, i.e., have a lower and upper bound that should
not be violated by the optimization algorithm. Let $\pulse_{\rmmin} \leq
\pulse(t) \leq \pulse_{\rmmax}$ be bounded. In order to restrict $\pulse(t)$ to
its bounds and to avoid manual truncation in case of violation of the bounds, we
internally parametrize $\pulse(t)$ via
\begin{subequations}
  \begin{align}
    u(t)
    &=
    \mathrm{arctanh}\left(
      \frac{%
        2 \pulse(t) - \pulse_{\rmmax} - \pulse_{\rmmin}
      }{%
        \pulse_{\rmmax} - \pulse_{\rmmin}
      }
    \right),
    \\
    \pulse(t)
    &=
    \frac{\pulse_{\rmmax} - \pulse_{\rmmin}}{2} \tanh\left(u(t)\right)
    +
    \frac{\pulse_{\rmmax} + \pulse_{\rmmin}}{2}.
  \end{align}
\end{subequations}
The auxiliary field $-\infty < u(t) < \infty$ is the one optimized in practice
and can be optimized boundless since, by construction, it can never violate the
boundaries of the actual field $\pulse(t)$ which it encodes.


%

\end{document}